\documentclass[apj]{emulateapj}
\usepackage{multirow,amsmath,longtable,appendix}

\usepackage{color}
\definecolor{lightgray}{gray}{0.8}

\received{16 November, 2014}
\revised{16 February, 2015 and 30 August, 2016.}

\shorttitle{Massive black holes}
\shortauthors{Graham, A.W.}
\begin{document}
\title{
Galaxy bulges and their massive black holes: a review\\
\vspace{3mm} 
{\it To appear in ``Galactic Bulges'', E. Laurikainen, R.F. Peletier, D.A
  Gadotti (eds.), Springer Publishing}
}
\author{Alister W.\ Graham\altaffilmark{1}}
\affil{Centre for Astrophysics and Supercomputing, Swinburne University
of Technology, Hawthorn, Victoria 3122, Australia.}
\altaffiltext{1}{Email: AGraham@swin.edu.au}

\begin{abstract}

With references to both key and oft-forgotten pioneering works,
  this article starts by presenting a review into how we came to believe in
  the existence of massive black holes at the centres of galaxies.  It then
  presents the historical development of the near-linear (black hole)--(host
  spheroid) mass relation, before explaining why this has recently been
  dramatically revised.  Past disagreement over the slope of the (black
  hole)--(velocity dispersion) relation is also explained, and the discovery
  of sub-structure within the (black hole)--(velocity dispersion) diagram is 
  discussed.  As the search for the fundamental connection between massive
  black holes and their host galaxies continues, the competing 
  array of additional black hole mass scaling relations for samples of 
  predominantly inactive galaxies are presented.
\end{abstract} 

\keywords{
 black hole physics -- galaxies: bulges -- galaxies: nuclei --
  galaxies: fundamental parameters}

\section{Overview}\label{sec:1}

Arguably one of the most exciting aspects of galaxy bulges are the monstrous
black holes which reside in their cores, sometimes lurking quietly, other
times beaming out their existence to the Universe.  Not only are they the
dominant species on the mass spectrum of individual objects, but they play
host to such a range of
extremely unusual phenomenon that they appeal to people of all ages and
professions.

For extragalactic astronomers, one curious aspect is the apparent coupling
between the mass of the black hole, $M_{\rm bh}$, and the host galaxy
bulge or spheroid, $M_{\rm sph}$, within which it resides.  The importance of
this is because it suggests that the growth of the two is intimately
intertwined, and unravelling this connection will provide insight into their
co-evolution.  While the $M_{\rm bh}$--$M_{\rm sph}$ relation may arise from
black hole feedback processes such that the black hole regulates the growth of
the surrounding spheroid (a remarkable feat given the factor of a billion difference in
physical size), correlations between both the central radial concentration of stars and
the central stellar density of the spheroid with $M_{\rm sph}$ might be telling us that 
it is instead the spheroid mass which (indirectly) dictates the black hole
mass through these relations. 

This article starts by providing a background briefing to the development of
ideas (since Einstein introduced his theories of relativity) which have led to
our current understanding of supermassive black holes in galactic nuclei
(Section~\ref{Sec_History}), and the eventual observational proof which ruled
out alternative astrophysical suggestions for the dark mass concentrations
identified there (Section~\ref{Sec_firm}).  Some 
effort has been made to reference key papers and give credit to the original
developers of ideas and solutions, of whom many have been poorly cited in the
literature to date. 

Not surprisingly, many reviews have been written about supermassive black
holes, and far more than the author was aware when approached to write
this review.  Enjoyable reports are provided by Kormendy \& Richstone (1995), 
Longair (1996 and 2006) which includes a well-written historical perspective, 
and an impressively extensive overview of many sub-topics can be found in 
Ferrarese \& Ford (2005) which remain highly relevant today. 
In it, they too provide an historical account of active galactic nuclei (AGN), 
detail the many methods used to measure the masses of black holes today, 
and compare the demographics of black holes in distant quasars with local
galaxies.  It is however their section~9, pertaining to the scaling relations
between the masses of black holes and the properties of their host galaxy 
that is the main focus of this article.  For references to other aspects of
massive black holes, over the past decade or so 
the following astrophysical reviews have focussed on: 
Sagittarius A$^*$ (Alexander 2005; Genzel et al.\ 2010); 
intermediate mass black holes (Miller \& Colbert 2004; van der Marel 2004); 
massive black hole binaries (Merritt \& Milosavljevi\'c 2005); 
AGN activity and feedback (Brandt \& Hasinger 2005; Ho 2008; McNamara \&
Nulsen 2007; Heckman \& Best 2014), including hot accretion flows (Yuan \& Narayan 2014) and cold
accretion flows (Kato 2008; Abramowicz \& Fragile 2013); 
connections with distant AGN (Shankar 2009a); 
redshifted fluorescent iron lines (Reynolds \& Nowak 2003; Miller 2007); 
gravitational radiation (Berti et al.\ 2009, see also Amaro-Seoane et al.\ 2012); 
black hole spin (Gammie et al.\ 2004; Reynolds 2013); 
black hole seeds (Volonteri 2010, see also Koushiappas et al.\ 2004); 
and a healthy mix of various topics (e.g.\ Kormendy \& Ho 2013; Genzel 2014)
as in Ferrarese \& Ford (2005). 

As noted by Ferrarese \& Ford (2005), in 2004 direct black hole mass
measurements were known for 30 galaxies, plus another 8 galaxies for which the
dynamical models might be in error.  Recently, Savorgnan \& Graham (2015), see also
Kormendy \& Ho (2013), tabulate 89 galaxies with reliably measured black hole
masses.  Not only has the sample size therefore tripled over the past decade,
but new scaling relations have been uncovered and old relations have been
revised --- and dramatically so as we shall see in the case of the $M_{\rm
  bh}$--$M_{\rm sph}$ and $M_{\rm bh}$--$L_{\rm sph}$ relations
(Section~\ref{Sec_M-M}).  The $M_{\rm bh}$--$\sigma$ relation, involving the
velocity dispersion of the galactic host, is reviewed in
Section~\ref{Sec_M-sigma} and the controversial issue of its slope addressed.
The apparent substructure in the $M_{\rm bh}$--$\sigma$ diagram, reported in
2008 due to barred galaxies and/or pseudobulges, is additionally discussed.

Having dealt in some detail with the two most commonly cited black hole
scaling relations in Sections~\ref{Sec_M-M} and \ref{Sec_M-sigma}, the
assortment of related relations are presented.  While not as popular in the
literature, it may be one of these relations which provides the fundamental,
or at least an important, link between the black hole mass and its host galaxy
(an issue raised by Alexander \& Hickox 2012).  Therefore, Section~\ref{Sec_n}
examines the connection between the black hole mass and the host spheroid's
S\'ersic index, i.e.\ how radially concentrated the spheroid's stellar
distribution is; this dictates the radial gradient of the gravitational
potential.  Section~\ref{Sec_mu0} describes the expected association between
the black hole mass and the central stellar density (prior to core depletion).
Section~\ref{Sec_core} explores the link between the mass of the black hole
and the missing stellar mass at the centres of giant spheroids.  The
connection between the black holes and the dense star clusters found in the
nuclei of many galaxies --- some of which may harbour intermediate mass black
holes --- is presented in Section~\ref{Sec_bh-nc}.  Section~\ref{Sec_halo}
discusses the black hole mass relation with the halo (baryons plus dark
matter) mass, expected to exist for spheroid dominated galaxies, while
Section~\ref{Sec_pitch} remarks on the existence of a correlation with the
pitch angle of spiral arms in late-type galaxies.  Finally,
Section~\ref{Sec_fun} considers the possibility that a third parameter may
account for some of the scatter in the above bivariate distributions, leading
to a more fundamental plane or hypersurface in 3-parameter space involving
black hole mass and two galaxy/spheroid parameters.

\section{Historical development: from mathematical speculation to widespread suspicion}
\label{Sec_History}

Karl Schwarzschild (1916, 1999; see also Droste 1917 who independently derived
the same solution in 1916) is widely recognised
for having developed the `Schwarzschild metric' for a spherical or point mass
within Einstein's (1916) theory of general relativity\footnote{It is of
  interest to note that Einstein was not keen on the idea of singularities,
  and in Einstein (1939) he wrote that {\it ``The essential result of this 
investigation is a clear understanding as to why the `Schwarzschild
singularities' do not exist in physical reality''}.}, but it was Finkelstein
(1958, see also Kruskal 1960) who realised the true nature of what has come to
be called the ``event horizon'' bounding these gravitational prisons.
Finkelstein eloquently describes this Schwarzschild surface as ``a perfect
unidirectional membrane: causal influences can cross it but only in one
direction''.  Five years later, while working at the University of Texas, the
New Zealand mathematician Roy Kerr (1963) formulated the metric for the more
realistic\footnote{Collapsing stars, and (accretion disc)-fed black holes, are
  expected to have substantial angular momentum.} {\it rotating} black hole.
Interestingly, solutions to this space-time include closed time-like curves
which, in theory, allow one to travel backwards in time (a concept popularised
but also questioned by Thorne 1994).  Kurt G\"odel (1949)
was actually the first to derive such strange solutions to the equations of
general relativity, although it is commonly suspected that all closed time-like
curves are just a mathematical artifact, in the same way that the original
singularity at the Schwarzschild radius was later explained away by a
coordinate transformation (e.g.\ Eddington 1924; Georges Lema\^itre 1933),
leaving just the singularity (i.e.\ black hole) at the centre.  
But even if we are to be denied our time machines\footnote{Time-travel enthusiasts might
appreciate a nod to the hypothetical Einstein-Rosen (1935) bridge (aka ``wormhole'', 
a term introduced by John Wheeler in 1957, e.g.\ Misner \& Wheeler
1957, and Klauder \& Wheeler 1957) 
which are warped regions of space-time within general relativity (Morris \& 
Thorne 1988; Morris et al.\ 1988; Hawking 1988). There is additionally the cosmic string time machine of Gott
(1991).}, black holes still offer the curious and unsuspecting property of
evaporating over time --- radiating like a black body --- before possibly then
exploding (Hawkings 1974, 1975).

Evolving parallel to the above analytical developments, our acceptance of
black holes as more than just a mathematical curiosity had additional
connections with stellar evolution and dark stars\footnote{John Michell (1783)
  was the first to calculate the existence of black holes, which he termed
  ``dark stars'', whose gravity was so strong that light would not be able to
  escape from their surface (see McCormmach 1968 and Schaffer 1979).
  Interestingly, Herschel (1791) subsequently speculated that `nebul{\ae}' might be
  regions of space where gravitationally-retarded particles of light are 
  endeavouring to fly off into space.}. 
As detailed by Yakovlev (1994), the Soviet physicist Yakov Frenkel (1928) was the
first to derive equations for the energy density and pressure of super-dense
stars comprised of a degenerate Fermi-gas of electrons of arbitrary
relativistic extent.  He is, however, not widely recognised for having done
so. Also using results from Albert Einstein's (1905) theory of special
relativity, Soviet physicist Wilhelm Anderson (1929) was the first to derive a
maximum mass for the fermion degenerate stellar model of white dwarf stars, 
above which the Fermi pressure is insufficient to overcome
gravity.  It is however the British physicist
Edmund Stoner (1929) who is somewhat better known for having presented the
structure for the mass, radius and density of white dwarf stars composed of
non-relativistic electrons.  Using his uniformly distributed mass density
model, Stoner (1930, see also Stoner 1932a,b) refined his work by formulating
how the core becomes relativistic at sufficiently high densities (as had
already been done by Frenkel 1928) and he too predicted a maximum stable mass 
(similar to Anderson 1929) for earth-sized, white dwarf stars. 
But it is Chandrasekhar (1931a, see also Chandrasekhar 1931b) who is well
known for calculating, in a short two-page article using {\it
  polytropic density models}, that at masses above $\approx$0.91 $M_{\odot}$,
electron-degenerate white dwarf stars are not stable.  That is, there is a
maximum mass (recognised today as 1.4 $M_{\odot}$) that white dwarf stars can
have.  If more massive than this limit then they must undergo further gravitational compression.
Soon after, Soviet physicist Lev Landau (1932) correctly identified that the
next level of resistance to their gravitational collapse 
would be met in the form of the denser neutron star 
(see also Oppenheimer \& Serber 1938; Oppenheimer \& Volkoff 1939).  Landau
(1932) and Chandrasekhar (1932, 1935)\footnote{Miller (2005) details the early
  work of Chandrasekhar on this topic.} predicted that the ultimate fate of an
evolved massive star would be to collapse to a singularity of infinite
density\footnote{In passing, it is noted that quark stars (Ivanenko \&
  Kurdgelaidze 1965) are also expected to have a stable configuration, en route
  between neutron stars and black holes.}.
Following further work on this idea (e.g.\ Baade \& Zwicky 1934; Zwicky 1938; Datt 1938), 
Oppenheimer \& Snyder (1939) carefully detailed how overly massive neutron stars 
are not stable and will collapse into stellar mass 
black holes.  Quite simply, if a star is massive enough and the outward
pressure from fusion is over, gravity will win over (e.g.\ Arnett 1967).

Wheeler (1966) wrote {\it ``In all the physics of the postwar era it is
  difficult to name any situation more enveloped in paradox than the
  phenomenon of gravitational collapse''}.  Then, in the following year
(1967), more than three decades after the initial prediction of neutron stars,
pulsars were discovered, finally signalling the existence of neutron stars
(Hewish et al.\ 1968; Pilkington et al.\ 1968; Hewish 1970).  Not
surprisingly, this bolstered belief in the existence of stellar mass black
holes (e.g.\ Penrose 1965; Vishveshwara 1970), as did 
(i) mathematical proof that a singularity will form if an event horizon has
formed (Penrose 1969; Hawking \& Penrose 1970), 
(ii) the X-ray pulses from Cygnus X-1 (Oda et al.\ 1971; Thorne \& Price 1975), 
and likely also (iii) the pioneering 
searches by Weber (1969, 1970) for gravitational radiation coming from 
even more massive objects at the centre of our Galaxy.

As detailed by Longair (1996, 2006, 2010), 
Ferrarese \& Ford (2005) and Collin (2006), the notion that the centres 
of galaxies may contain massive black holes, millions to hundreds of millions
times the mass of our Sun, stems from the discovery of the great distance to,
and thus luminosity of, the quasi-stellar radio source 3C~273.  The optical
counterpart of this radio source was cleverly discovered by Hazard, Mackey \&
Shimmins (1963) using the Parkes radio telescope and lunar eclipsing. Its 
redshift was subsequently taken with the Palomar Observatory's Hale telescope and correctly
interpreted by Schmidt (1963), see also Oke (1963) regarding 3C~273 and
Greenstein \& Matthews (1963a,b) in the case of 3C~48 (whose redshift had
remained uninterrupted over the preceding couple of years).
 
Baade \& Minkowski (1954), Ambartsumian (1958), Woltjer (1959), 
Burbidge (1959), Burbidge et al.\ (1963, 1964), Lynds \& Sandage (1963) and others 
had already recognised active galactic nuclei (AGN) to be incredibly energetic 
phenomena.\footnote{While relatively low-luminosity Seyfert (1943) 
galaxies --- with broad emission lines as previously observed 
by Fath (1909) and Slipher (1917) ---  
were of course already known in 1963, it 
was not yet fully appreciated that quasars are their high-energy kin, 
although similarities were noted by Burbidge et al.\ (1963) and Burbidge
(1964).} 
Radio galaxy 3C~273 and other active galactic nuclei emit vasts amount of
energy from a small volume of space (as indicated by quasar variability on
short time scales, Smith \& Hoffleit 1963)\footnote{Reviews of AGN and their
  variability are given by Mushotzky et al.\ (1993), Ulrich et
  al.\ (1997), and Peterson (1997).}  and were thus thought to be powered from
the gravitational potential energy\footnote{As stated by Rees (1998), the
  black hole's gravitational well ``must be deep enough to allow several percent of
  the rest mass of infalling material to be converted into kinetic energy, and
  then radiated away from a region compact enough to vary on timescales as
  short as an hour.''}  released as matter falls onto a compact massive object
(Salpeter 1964; Zel'dovich 1964; Zel'dovich \& Novikov 1964; Ne'eman
1965; Shakura \& Sunyaev 1973)\footnote{Like many capable theorists, Zel'dovich and Novikov did not
  restrict themselves to one theory, and in Bisnovatyi-Kogan, Zel'dovich \&
  Novikov (1967) they proposed that quasars may be billion solar mass stars
  burning brightly for tens of thousands of years, see also Hoyle \& Fowler
  (1963), while Novikov (1965) additionally advocated what we now know as `white holes'.}.
Based upon Eddington-limiting arguments at the time, it was immediately
realised that the central object has to be massive or else the radiation
pressure of the quasar would literally blow the quasar apart.  Hoyle et
al.\ (1964) acknowledged the possibility of ``invisible mass'' perhaps from
imploded objects of very large mass\footnote{Hoyle \& Burbidge (1966) also
  speculated that quasars may be nearby objects and that their redshifts do
  not necessarily reflect the expansion of the universe, see also Hoyle et al.\ (2000) and
  Burbidge et al.\ (2006).}.
 
Just two years after the high-redshifts were recorded for the
star-like\footnote{Faint halos had been reported around some of these
  `star-like' objects, which we now know is due to the host galaxy surrounding
  the bright AGN (e.g.\  Gehren et al.\ 1984; Hutchings et al.\ 1984, and 
references therein).} 
radio sources, Sandage (1965) reported on the high
abundance of radio-quiet quasars, referring to them as a ``major new
constituent of the universe''.  What he had revealed was that in addition to
the radio-loud quasars, the Universe was teeming with many more quasars.
Encapsulating the ideas of recent years, Lynden-Bell\footnote{Historical
  footnote: The daily commute along the A273 to Herstmonceux in Sussex
  prompted Donald Lynden-Bell to find a satisfactory explanation for the 
  quasar 3C~273 (priv.\ comm.\ 2015).} (1969) and Lynden-Bell \&
Ress (1971) suggested that a massive black hole resides at the cores of many galaxies
(see also Wolfe \& Burbidge 1970), and that the infall of
orbital matter builds an accretion disc (e.g.\ Thorne 1974) 
which heats up due to friction.  For a rapidly spinning 
black hole, this process can liberate a substantial fraction (up to 0.42 for a
maximally spinning black hole) of the 
infalling matter's rest mass energy\footnote{For 
  comparison, nuclear fusion is known to release less than 1\% of the rest 
  mass energy (0.7\% in the conversion of hydrogen to helium) and thus
  `super-stars' are not as efficient sources of energy as rapidly spinning
  accretion discs around supermassive black holes.} 
(Bardeen \& Wagoner 1969; Bardeen 1970). 
Further support for the presence of massive black holes were the linear radio
features emanating from the nuclei of galaxies --- which were likely emitted
from a stable gyroscope such as a spinning black hole --- and the 
superluminal speed of these radio jets (e.g.\ Cohen et al.\ 1971; Whitney et
al.\ 1971). 

The moniker ``black hole'' was used by Ann Ewing (1964) 
just a year after the redshift of 3C~273 was announced.
She reportedly heard it at a meeting of the American Association for the 
Advancement of Science (AAAS), and it was later seen used in a scientific paper
by John Wheeler
(1968)\footnote{John Wheeler is first recorded to have used the 
term ``black hole'' at his 27 December, 1967, AAAS invited lecture, a few
years after Ann Ewing. 
However, given that he coined the term ``worm hole'', it seems likely that 
he also introduced the expression ``black hole'', although the author does not
rule out that it may have been Fritz Zwicky.}.  
By 1970 the label appears as a familiar term in the literature.  
It had, at times, previously been used to describe 
dusty dark patches in our own Galaxy (e.g.\ Barnard 1897; Campbell 1917). 
However it became the popular replacement for what the Soviet physicists
(e.g.\ Zel’dovich 1964) called a frozen star\footnote{For an external
  observer, time appears to stop inside the Schwarzschild radius, giving rise to the
  term ``frozen star'', because the collapse of a star will appear to freeze
  once the star is within the event horizon.}, and Western physicists 
called a collapsed star or a ``collapsar'' (e.g.\ Cameron 1971).  
The term ``singularity'' had also been, and still is, regularly used by the 
mathematicians to indicate where any quantity in the field equations becomes 
infinite.

\begin{figure*}
\includegraphics[scale=0.63, angle=-90]{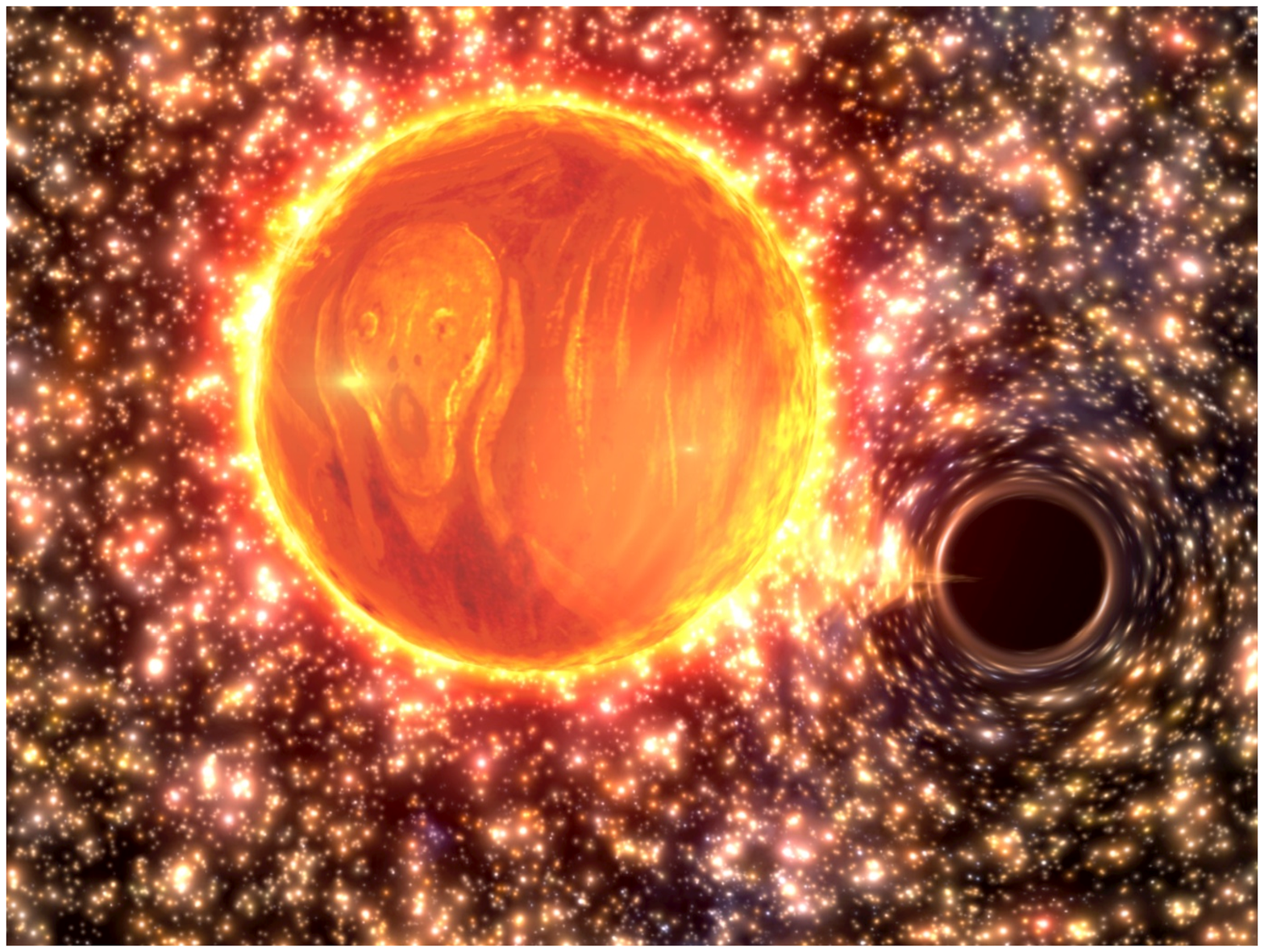}
\caption{Artist's impression of the horror at a galactic centre. 
Credit: Gabriel P\'erez D\'iaz. [To appear in the published version.]}
\label{Fig1}
\end{figure*}

Despite its strangely endearing name, 
the phrase ``black hole'' is often noted to be somewhat unfortunate in that it
implies a {\it hole} in space through which matter may fall through. 
The idea of an actual singularity --- a point of infinite density which 
arises out of classical physics after division by 0 --- is also not popular
and considered rather old-school.  
While a Planck-sized mote may be a better description, 
what actually exists near the centre of a black hole's event horizon is hotly
debated.  Mathematically-inclined readers who are interested in what 
a black hole may be like, might enjoy reading about the 
'fuzzball' picture from string theory ('t Hooft 1980; Mathur 2005, and
associated references), 
or descriptions of black holes in quantum gravity theories such as spin
foam networks (e.g.\ Penrose 1971a,b; Penrose \& Rindler 1986; 
Rovelli 1998; Domagala \& Lewandowski 2004; Perez 2004) 
or loop quantum gravity (e.g.\ Ashtekar \& Bojowald 2005; Hayward 2006).

\section{On firmer ground} 
\label{Sec_firm}

Acceptance of the idea that supermassive black holes reside at the centres of
galaxies was not as straight forward as suggested above.  During the 1960s and
1970s the AGN community battled it out amongst themselves before (largely)
embracing the idea that black holes must be required to power the quasar
engines of galaxies.  Building on Sandage (1965), Soltan (1982) reasoned that
there had to be a lot of mass locked up today in massive black holes because
of all the past quasar activity, and Rees (1984) advocated further for the
preponderance of massive black holes in the nuclei of galaxies.  Then during
the 1980s and early 1990s it was primarily the inactive-galaxy community, as
opposed to the AGN-community, who remained skeptical until two key papers in
1995 (discussed shortly).

Among the pioneering observational papers for the presence of a massive black
hole in individual, nearby, non-AGN galaxies, Sanders \& Lowinger (1972)
calculated that the Milky Way houses a $0.6\times10^6 M_{\odot}$ black hole
and Sargent et al.\ (1978) 
concluded that a $5\times10^9 M_{\odot}$ black hole
very probably exists in M87 (see also Lynden-Bell 1969 who predicted a
$30\times10^6 M_{\odot}$ black hole for the Milky Way\footnote{With the 
benefit of hindsight, we now know that radio synchrotron emission from 
Sagittarius~A was first seen in the 5 GHz data from Ekers \& Lynden-Bell (1971).}, and a $40\times10^9$
black hole in M87, i.e.\ an order of magnitude higher).  Although these
works had revealed that very high masses in small volumes were
required at the centres of these galaxies (see also Dressler 1984 and Tonry
1984 in the case of M31 and M32, respectively), 
it took some years before the observations / measurements improved and 
alternatives such as a dense cloud of stellar mass black holes or neutron
stars could be ruled out.  The three following observational works turned the
tide of opinion among the remaining naysayers who demanded further proof
before accepting the existence of what is indeed an extreme astrophysical
object: the supermassive black hole.

1) Before an object crosses within a black hole's event horizon, any radiation
it emits away from the black hole will be gravitationally redshifted, the
extent of which depending on how close the object is to the event horizon.
Such a tell-tale signature of redshifting was reported on 22 June 1995 by
Tanaka et al.\ (1995) who detected the highly broadened, ionised iron
K$\alpha$ line (6.4 keV) from the galaxy MCG-6-30-15.  This highly asymmetric,
predominantly redshifted, X-ray emission line had a width corresponding to
roughly one-third of the speed of light, and was thought to have been emitted
at just 3 to 10 Schwarzschild radii from the black hole.  Such relativistic
broadening has since been shown to be commonplace (Nandra et al.\ 1997),
thanks to the enhanced sensitivity and spectral resolution of the Japanese 
ASCA X-ray satellite (Tanaka et al.\ 1994). 

2) Additional convincing evidence for the reality of massive black holes had
came from the very high mass density required to explain the central object in
the Seyfert galaxy NGC~4258 (M106).  Using the Very Long Baseline Array in New
Mexico, Miyoshi et al.\ (1995) showed that the H$_2$O maser emission from this
galaxy originates from a thin, rotating nuclear gas disc/annulus displaying a
clear Keplerian rotation curve and requiring a mass of $3.6\times10^7
M_{\odot}$ within a size of just 0.13 parsec\footnote{Based on a galaxy
  distance $D = 6.4$ Mpc.}  (see also Haschick et al.\ 1994, Watson \& Wallin
1994, and Greenhill et al.\ 1995a,b).  In their January 12 paper, Miyoshi et
al.\ (1995) note that the short collisional timescale ($<10^8$ years) for a
swarm of solar mass dark stars with such density ($>4\times10^9 M_{\odot}$
pc$^{-3}$ inside of the inner 4.1 milliarcseconds) implies that such a
hypothetical star cluster could not survive (see also Maoz 1995, 1998); a
single supermassive black hole is the only viable candidate.  A second example
of extreme mass density ($3.2\pm0.9 \times 10^8 M_{\odot}$ pc$^{-3}$) has
since been shown in the Circinus galaxy by Greenhill et al.\ (2003).

3) Several years later, high spatial resolution measurements of stellar orbits
around the central object in our own Milky Way galaxy also eventually ruled
out the possibility that it could be a swarm of neutron stars or stellar mass
black holes, with the high density favouring the existence of a massive black
hole (Sch\"odel et al.\ 2002; Ghez et al.\ 2005, 2008; Gillessen et
al.\ 2009).  confirming earlier suspicions
(Lacy et al.\ 1979, 1980; Eckart \& Genzel 1996, 1997; Genzel et al.\ 1996,
1997; Ghez et al.\ 1998; see also Alexander 2005 and references therein).

As was appropriately emphasized by Merritt \& Ferrarese (2001b), within the
black hole's sphere-of-influence --- whose radius is defined as $r_{\rm infl}
= G M_{\rm bh} / \sigma_{\rm sph}^2$ (e.g.\ Peebles 1972; Frank \& Rees 1976)
where $\sigma_{\rm sph}$ is roughly the host spheroid's velocity dispersion
immediately beyond $r_{\rm infl}$ --- one expects to find Keplerian dynamics
which are dominated by the black hole.  The velocity dispersion of the stars
(or the rotational velocity of a relatively lighter disc, as in the case of
NGC~4258) inside $r_{\rm infl}$ should thus decline with the inverse square
root of the radius, i.e.\ $\sigma(R) \propto R^{-0.5}$, just as rotational
velocities of Keplerian discs or solar systems have $v_{\rm rot} \propto 1/
\sqrt{R}$.

The absence of this clear detection for many galaxies has led Merritt (2013)
to question their reported black hole measurements, which may be better
interpreted as upper limits until we are better able to resolve the
sphere-of-influence (see also Valluri, Merritt \& Emsellem 2004).  With this
cautionary note, we proceed to the topic of black hole scaling relations,
which at the very least would still be upper envelopes in the various diagrams
of black hole mass versus host spheroid properties.  It may however then be
unusual that all of the clear-cut examples for a definitive black hole reside
on this upper envelope (but see Ford et al.\ 1998; Ho 1999, his section~7; and
Batcheldor 2010).

\section{The $M_{\rm bh}$--$L_{\rm sph}$ and $M_{\rm bh}$--$M_{\rm sph}$ relations}
\label{Sec_M-M}

Commenting on the ratio of black hole mass to spheroid mass in M31 and M32,
Dressler \& Richstone (1988) suspected a relation, and used it to predict
billion solar mass black holes in bright elliptical galaxies. While the
prediction was not new, in the sense that the authors were aware that past
theoretical papers had stated that quasars in big elliptical galaxies could
have $10^9$ solar mass black holes (e.g.\ Rees 1984; Begelman, Blandford \&
Rees 1984, and references therein)\footnote,
the idea of a scaling relation with the {\it spheroid} does seem to be 
new\footnote{Jarvis \& Dubath (1988) appear to have also picked up on this connection 
when they wrote in regard to M31
and M32 that: ``The likely presence of black holes in two of the closest
  galaxies with bulge-like components compels us to look at the nuclei
  of other nearby or large galaxies'', which they 
did for the Sombrero galaxy (NGC~4594).}. 
Dressler (1989) further advocated this connection between the black hole and
the host spheroid (not the disc), and from a sample of 5 galaxies he noted
that there is a ``rough scaling of black hole mass with the mass of the spheroidal
component''.

This differed slightly from Hutchings et al.\ (1984) who had reported that the
``black hole [] mass is related to that of the galaxy, increasing 60\% faster
than that of the galaxy''.  The study by Hutchings et al.\ (1984) was of poorly resolved,
distant quasars which prevented them from performing a bulge/disc decomposition and
as such they did not report on a black hole mass relation with the host
spheroid.  However, there is an upper limit to the brightness of quasars which
has been observed to scale with the brightness of the host galaxy (which are
typically spheroid-dominated for the brightest quasars).  Using real data, Yee (1992) fit a
linear relation to this limit, which he called the $M_{\rm QSO}$--$M_{\rm G}$
relationship, and wrote that ``it may arise due to a correlation of the mass
of the central engine and the galaxy mass'', such that ``the brightest quasars
for a given galaxy mass are the ones shining at or near the Eddington limit
(which is set by the mass of the central engine), while others are at lower
luminosities''\footnote{In passing, and as noted by Alexander \& Natarajan (2014), 
it is possible to exceed the Eddington limit to black hole growth 
(as noted by Begelman 1979 and Soffel 1982), due to an effective gas drag of the photons.  
Moreover, non-spherical accretion in the form of a disc can also result in 
black hole growth superseding the Eddington limit (Nayakshin et al.\ 2012).}.
As noted by McLeod (1997, see also McLeod et al.\ 1999 and result number 4
from Laor et al.\ 1997), Yee (1992) had 
effectively discovered the linear, high-mass end of the $M_{\rm bh}$--$M_{\rm
  sph}$ distribution. 

With three more galaxies than Dressler (1989), Kormendy \& Richstone (1995,
see also Kormendy 1993) 
wrote a review article in which they plotted this data and reiterated in
mathematical form what Dressler had said, and Yee (1992) had shown for massive
bulges, 
i.e.\ $M_{\rm bh} \propto M_{\rm bulge}$.  While they did not fit a relation
to the data, they did report a mean $M_{\rm bh}/M_{\rm bulge}$ ratio of 0.22\%
(including the Milky Way) and thereby effectively created a more quantitative
basis for a linear $M_{\rm bh}$--$M_{\rm bulge}$ relation.

Following the prediction by Haehnelt \& Rees (1993) that $\approx$30\% of
nearby galaxies likely house a central massive black hole, Kormendy \&
Richstone (1995) remarked that at least 20\% of nearby galaxies possess such a
black hole -- while noting that alternatives such as massive concentrations of
dark stars could not yet be ruled out.  Magorrian et al.\ (1998) built on this
and suggested that {\it most} nearby galaxies harbour a massive black hole
(see also Sigurdsson \& Rees 1997, and the reviews by Ford et al.\ 1998 and
Richstone et al.\ 1998), supporting the strong suspicion held by many
(e.g.\ Blandford 1986; Rees 1990).  Moreover, this followed closely on the
heels of the observation that many quiescent galaxies have weak central radio
sources (e.g.\ Keel 1985; Sadler et al.\ 1989, 1995; Ho et al.\ 1997), likely
signalling low-level accretion onto near-dead quasars.

Rather than the pure `linear' scaling, a single power-law relation was
introduced by Magorrian et al.\ (1998; see also Franceschini et al.\ 1998) 
to describe the distribution of 32
points in the $M_{\rm bh}$--$M_{\rm bulge}$ diagram, such that the log-linear
slope was 0.96$\pm$0.12 (which is of course still consistent with a slope of 1
and thus a linear relation)\footnote{As suspected by Magorrian et al.\ (1998),
  and noted by van der Marel (1999) and Gebhardt et al.\ (2000), their use of
  a two-integral distribution function which ignores radial
  velocity-dispersion anisotropy (see Binney \& Mamon 1982) caused them to
  over-estimate the black hole masses by an average factor of 3--4.5}.  In
other works, using variously updated masses and samples, Ho (1999) reported a
median $M_{\rm bh}/M_{\rm bulge}$ ratio of 0.2\%, and Merritt \& Ferrarese
(2001c) and Kormendy \& Gebhardt (2001) reported a ratio of 0.13\%, although
with notable scatter.  McLure \& Dunlop (2002) noticed that the scatter was
considerably reduced once the disc galaxies were excluded, suggestive of poor
bulge/disc decompositions used to estimate the bulge masses.  Marconi \& Hunt (2003) subsequently performed
careful bulge/disc decompositions on near-infrared $K$-band images, less
effected by dust and star formation.  They also showed that the
dynamical/virial mass of the spheroid correlated linearly with the black hole
mass, and H\"aring \& Rix (2004) provided improved dynamical masses for the
derivation of their near-linear relation.  For the next decade, studies of the
$M_{\rm bh}$--$L_{\rm bulge}$ and $M_{\rm bh}$--$M_{\rm bulge}$ diagram
remained dominated by high-mass galaxies\footnote{Studies were also biased by
  the inclusion of one or two rare ``compact elliptical'' galaxies
  (e.g.\ M32 in Graham 2007b and G{\"u}ltekin et al.\ 2009, their Fig.4) that do not
  represent the population at large.} having $M_{\rm bh} >$ $\sim$$0.5\times10^8
M_{\odot}$ and, despite each paper's incremental improvements, continually
recovered a single, near-linear $M_{\rm bh}$--$M_{\rm bulge}$ relation
(e.g.\ Ferrarese \& Ford 2005; Lauer et al.\ 2007; Graham 2007b, 2008a, his
section~6; G{\"u}ltekin et al.\ 2009; Sani et 
al.\ 2011; Beifiori et al.\ 2012; Erwin \& Gadotti 2012; Vika et al.\ 2012;
van den Bosch et al.\ 2012; McConnell \& Ma 2013; Rusli
et al.\ 2013a).  A recent notable exception has been L\"asker et al.\ (2014b)
who advocate, with a near-infrared sample of 35 galaxies, that the black hole mass
correlates equally well with the total (bulge plus disc) luminosity as it does
with the bulge luminosity at 2.2 $\mu$m, and that one has 
$M_{\rm bh} \propto L^{0.75\pm0.10}_{\rm bulge}$ and 
$M_{\rm bh} \propto L^{0.92\pm0.14}_{\rm galaxy}$. They attribute this to
the smaller bulge fluxes obtained from their decomposition of the
galaxies' light and the type of linear regression performed. 
Savorgnan et al.\ (2016) have, however, since included 17, rather than 4, spiral
galaxies and found that it is indeed the bulge rather than galaxy mass which
has the strongest correlation.

There were a few early deviations from the above (near) 
convergence of opinion on a linear relation that should be noted. 
First, while the Abstract of Laor (1998) largely supports the linear
relation of Magorrian et al.\ (1998), the main text reports that $M_{\rm bh} \propto
M^{1.5-1.8}_{bulge}$ (although it suggests that this may be partly due
to the fact that all their lower mass quasar hosts are disc galaxies for which
they may have over-estimated the bulge mass) and 
Second, it also notes that 
the low-mass inactive galaxies from Magorrian et al.\ (1998) better match
their steeper $M_{\rm bh}$--$M_{\rm bulge}$ relation than the linear one. 
Third, Wandel (1999) reported a mean $\log (M_{\rm bh}/M_{\rm bulge})$ ratio
of $-3.5$ for a sample of Seyfert galaxies with black hole masses
predominantly less than $10^8 \, M_{\odot}$.  This is 0.6 dex, i.e.\ a factor
of 4, smaller than reported by Merritt \& Ferrarese (2001c) and Kormendy \&
Gebhardt (2001) who used a sample with $\sim$80\% of the galaxies having
$M_{\rm bh} > 0.8\times 10^8 \, M_{\odot}$. Wandel (1999) argued and wrote
{\it ``It is plausible, therefore, that the Seyfert galaxies in our sample
  represent a larger population of galaxies with low BBRs [black hole to bulge
    mass ratios], which is underrepresented in the Magorrian et
  al.\ sample''}\footnote{McLure \& Dunlop (2001) correctly noted that a
  better bulge/disc decomposition reduces the observed flux attributed to the
  bulges by Wandel (1999), however the dust corrections which were not applied can largely
  cancel this reduction (compare figures~1 and 7 in Graham \& Worley 2008).}.

Fourth, 
while Wandel reported $M_{\rm bh} \propto L^{1.4}_{\rm bulge}$ (which equates
to $M_{\rm bh} \propto M^{1.2}_{\rm bulge}$ when using the same $M/L \propto
L^{0.18}$ relation as Laor 1998 and Magorrian et al.\ 1998), 
the data in Wandel (1999, their figure~1) reveal that a relation with
a slope steeper than 1.4 would be likely from a {\it symmetrical} regression. 
Fifth, using upper limits for black hole masses, Salucci et al.\ (2000)
reported on hints that the $M_{\rm bh}$--$M_{\rm bulge}$ relation is 
significantly steeper in spiral galaxies than in [massive] elliptical galaxies. 
Finally, Laor (2001) reinforced his claim that a steeper, single power-law seems more
applicable than a linear relation, finding $M_{\rm bh} \propto 
M^{1.53\pm0.14}_{\rm bulge}$. 
Related to this, 
Ryan et al.\ (2007) further reveals that the linear $M_{\rm bh}$--$M_{\rm
  bulge}$ relation over-estimates the masses of black holes in low-mass
Seyfert galaxies.

\subsection{A bend in the road}
\label{SubSec_bend}

Before beginning this section, it is necessary to introduce some nomenclature
which may be unfamiliar to some readers.  The term ``S\'ersic galaxy'' or
``S\'ersic spheroid'' shall be used to denote galaxies or spheroids (elliptical
galaxies and the bulges of disc galaxies) whose surface brightness profile is
well described by the S\'ersic (1963, 1968) model all the way into the centre
of the galaxy.  Two decades ago Caon et al.\ (1993) demonstrated that the S\'ersic
model fits the surface brightness profiles of early-type galaxies remarkably
well over a large dynamic range.  An historical and modern review of
S\'ersic's model can be found in Graham \& Driver (2005). S\'ersic galaxies may contain
additional nuclear flux components above that of the host S\'ersic spheroid.  The term
``core-S\'ersic galaxy'' or ``core-S\'ersic spheroid'' refers to a galaxy whose
main spheroidal component has a partially-depleted core (i.e.\ a central
stellar deficit of light that is not due to 
dust) such that the surface brightness profile is well described by the
core-S\'ersic model (Graham et al.\ 2003b).  The history of galaxy surface
brightness models and the impact that the above systematically (with
luminosity) varying structures (i.e.\ non-homology and depleted cores) have on
galaxy scaling laws and the unification of bright and faint early-type
S\'ersic galaxies is discussed at length in Graham (2013).

\vspace{4mm}

Re-analysing the dynamical spheroid mass and (updated) black hole mass data
for 30 galaxies studied by H\"aring \& Rix (2004), but this time separating the
galaxies depending on whether or not they have a partially depleted core,
Graham (2012a) found that the two populations follow different relations in the
$M_{\rm bh}$--$M_{\rm sph,dyn}$ diagram.  While the dozen core-S\'ersic spheroids, 
which are the more massive spheroids, followed the near-linear relation
$M_{\rm bh} \propto M^{1.01\pm0.52}_{\rm sph},dyn$, the S\'ersic spheroids
followed a much steeper power-law relation, such that $M_{\rm bh} \propto
M^{2.30\pm0.47}_{\rm sph,dyn}$.  Excluding the barred galaxies, the S\'ersic 
relation was $M_{\rm bh} \propto M^{1.92\pm0.38}_{\rm sph,dyn}$.  This
near-quadratic relation for the low- and intermediate-mass spheroids
had never been reported before and it signalled
a bend in the $M_{\rm bh}$--$M_{\rm sph,dyn}$ diagram.  

With an increased sample size of 72 galaxies with directly measured black hole
masses, Graham \& Scott (2013) confirmed this behavior using near-infrared
$K_s$-band magnitudes.  Their sample of two dozen core-S\'ersic spheroids gave
$M_{\rm bh} \propto L^{1.10\pm0.20}_{\rm sph}$, while the four dozen S\'ersic
spheroids gave the relationship $M_{\rm bh} \propto L^{2.73\pm0.55}_{\rm
 sph}$, which reduced to 
$M_{\rm bh} \propto M^{2.34\pm0.47}_{\rm sph,dyn}$
when using $M_{\rm dyn}/L_K \propto L_K^{1/6}$ (e.g., Magoulas et al.\ 2012;
La Barbera et al.\ 2010).  Employing the {\sc archangel} photometry pipeline
(Schombert \& Smith 2012) applied to Two Micron All-Sky Survey images
(Skrutskie et al.\ 2006), which effectively corrects for missing light at
large radii, Scott et al.\ (2013) converted the $K_s$-band magnitudes of the
spheroids into stellar masses.  They found that $M_{\rm bh} \propto
M^{0.97\pm0.14}_{\rm sph,*}$ and $M_{\rm bh} \propto M^{2.22\pm0.58}_{\rm
  sph,*}$ for the S\'ersic spheroids and core-S\'ersic, respectively.

We therefore now have a situation which is dramatically different to what was
believed for the past two decades.  It is not simply that we no longer have a
single, near-linear $M_{\rm bh}$--$M_{\rm sph}$ relation for all spheroids,
but the main growth phase of black holes and bulges, involving gas rich
processes, follows a near-quadratic relation, with gas-poor ``dry'' mergers
subsequently creating the core-S\'ersic galaxies which depart from the
high-mass end of this near-quadratic relation\footnote{Some S\'ersic galaxies
  may follow the near-linear $M_{\rm bh}$--$M_{\rm sph}$ relation, having
  experienced a major dry merger event in which the nuclear star clusters from
  the progenitor galaxies have been eroded away but an obvious partially
  depleted core is not yet formed (see Bekki \& Graham 2010).  These may well
  be the galaxies at $-19.5 > M_B > -20$ mag in C\^ot\'e et al.\ (2007. their
  figure~3e).\label{foot_Bekki}}.  That is, the growth of massive black holes
has been much more rapid than that of their host spheroids.  

Naturally, the
simple addition of galaxies and their black holes, through dry merging, will
establish the observed near-linear relation for the core-S\'ersic galaxies.
The average $M_{\rm bh}/M_{\rm sph}$ ratio of these core-S\'ersic galaxies
then reflects the value obtained at the high-mass end of the near-quadratic
S\'ersic $M_{\rm bh}$--$M_{\rm sph}$ relation from which they peeled off.  
In late 2012 Graham \& Scott (2013) reported this mass 
ratio to be 0.49\%, in agreement with that already noted by Laor (2001) for
massive spheroids.  This ratio is basically the calibration for the Yee (1992) relation
between black hole mass and galaxy mass in massive galaxies, modulo the fact
that some core-S\'ersic galaxies contain large discs.  Furthermore, our own
galaxy, with an $M_{\rm bh}/M_{\rm sph}$ ratio of 0.05\%, is no longer a low
outlying point requiring explanation in the $M_{\rm bh}$--$M_{\rm sph}$
diagram.  It has a mass ratio in accord with the near-quadratic scaling
relation for S\'ersic spheroids.

\begin{figure}[t]
\includegraphics[scale=.43,angle=-90]{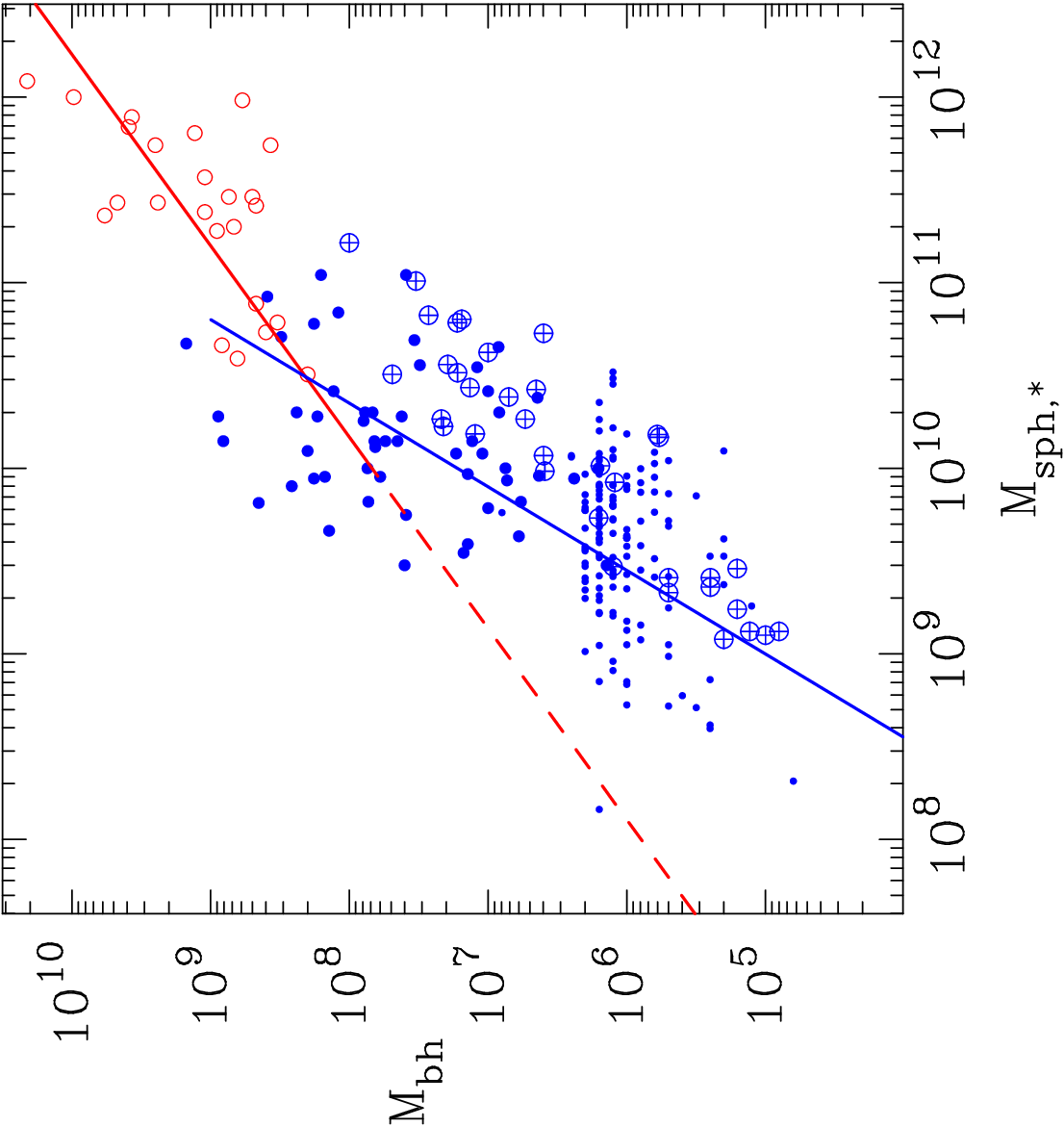}
\caption{Black hole mass versus host spheroid's stellar mass (in units of solar mass).
Core-S\'ersic spheroids are shown with open red circles, while S\'ersic
spheroids are shown by the large blue dots.  A sample of 139 low mass AGN
from Jiang et al.\ (2011) are denoted by the small dots, while an additional
35 higher mass AGN (which may have had their host spheroid masses over-estimated 
by overly-high $(M/L)_{\rm stellar}$ 
ratios, see Busch et al.\ 2014) are denoted by the cross hairs. 
The optimal near-linear and near-quadratic scaling relations from Scott et al.\ (2013)
are shown as the red (solid and dashed) and blue (solid) line for the
core-S\'ersic and S\'ersic spheroids, respectively.  
Of note is that 
68\% of the 139 AGN (i.e.\ +/-34\%) are contained within 0.83 dex in the
horizontal direction, representing a level of scatter equal to that about the 
near-linear relation observed at the high-mass end. 
The non-AGN S\'ersic galaxies have more scatter than the non-AGN core-S\'ersic
galaxies because of the crude way in which their bulge masses were estimated
(see Graham \& Scott 2015, from which this figure is taken). 
}
\label{Fig2}
\end{figure}

Adding AGN data from half a dozen recent papers which had observed the AGN
black hole masses to reside below the original $M_{\rm bh}$--$M_{\rm sph}$
relation, Graham \& Scott (2015) revealed that they depart from the
near-linear $M_{\rm bh}$--$M_{\rm sph}$
relation in a systematic manner consistent with the near-quadratic $M_{\rm
  bh}$--$M_{\rm sph}$ mass scaling relation for S\'ersic galaxies.
That is, they are not randomly offset.  This is shown in
Figure~\ref{Fig2}.  This also provides the picture with which we can now
interpret the observations by Laor (1998, 2001) and Wandel (1999), who were on
the right track over a decade ago. 

{\it If} one was to separate the galaxies in
Figure~\ref{Fig2} at $M_{\rm bh} = 2 \times 10^6 \, M_{\odot}$, one would
(understandably but inappropriately) conclude that the lower mass spheroids do
not follow an $M_{\rm bh}$--$M_{\rm sph,*}$ relation (Jiang et al.\ 2011).
This had resulted in these lower mass spheroids being considered distinct by
some, and sometimes labelled `pseudobulges' as opposed to `classical' bulges
(Gadotti \& Kauffmann 2009; Kormendy, Bender \& Cornell 2011) with the
separation said to occur at $n=2$. This is also where the alleged divide
between dwarf elliptical and ordinary elliptical galaxies was said to occur
($M_B = -18$ mag, $M_{\rm gal,*} \approx 2\times 10^{10} \, M_{\odot}$ $n
\approx$ 2--2.5, $\sigma \approx$ 100--120 km s$^{-1}$).  However, without the
fuller parameter baseline that we now have, or artificially
subdividing the data at a S\'ersic index of 2, or at $M_B = -18$ mag, or where
the curvature in relations using `effective' radii and surface brightnesses
are a maximum (see Graham 2013 for an explanation of this), the continuity
between the low- and intermediate-luminosity S\'ersic galaxies can be missed,
even if the data itself is accurate.  This issue is discussed further in
section~\ref{Sec_pseudo}. 

The distribution of points in Figure~\ref{Fig2} reveals that black holes grow 
faster than the stellar population of their host spheroids, for which abundant
evidence is now appearing (e.g.\ Diamond-Stanic \& Rieke 2012; Seymour et
al.\ 2012; Trakhtenbrot \& Netzer 2012; Agarwal et al.\ 2013; Alonso-Herrero et
al.\ 2013; LaMassa et al.\ 2013 Lehmer et al.\ 2013; Drouart et al.\ 2014). 
For example, Diamond-Stanic \& Rieke (2012) report that the black hole growth
rate is proportional to the 1.67 (=1/0.6) power of the star formation rate
within the inner kpc (roughly the bulge half-light radii) of their Seyfert
galaxies, while the analysis from LaMassa et al.\ (2013) gives an exponent of
2.78 (=1/0.36) for their sample of $\sim$28,000 obscured active galaxies, 
quite different from the linear value of 1. 

Figure~\ref{Fig2} also reveals that classical bulges, pseudobulges,
clump-bulges (Noguchi 1999), and mixed-bulges containing both a classical
bulge and a pseudobulge, all follow the steeper scaling relation, until the
onset of relatively dry mergers revealed by the scoured cores seen in the 
centres of (many of) the most massive spheroids.

With their supernova feedback producing a steeper relation than their AGN
feedback prescription, the models of Cirasuolo et al.\ (2005, their Figure~5)
show a bend in the $M_{\rm bh}$--$M_{\rm sph}$ (and $M_{\rm
  bh}$--$M_{\sigma}$) relation at $M_{\rm bh}\approx 10^8 M_{\odot}$.  At
these lower masses, a steeper than linear $M_{\rm bh}$--$M_{\rm sph}$ relation
can also be seen in the differing models of Dubois et al.\ (2012), Khandai et 
al.\ (2012, their Figure 7); Bonoli et al.\ (2014, their Figure 7) and
Neistein \& Netzer (2014, their Figure~8).

What happens in the $M_{\rm bh}$--$M_{\rm sph}$ diagram at black hole masses
less than $10^5 \, M_{\odot}$ is not yet known, although LEDA~87300 suggests
that the steep relation continues (Graham, Ciambur \& Soria 2016).  While the absence of a
definitive black hole detection in M33 (Kormendy \& McClure 1993; Gebhardt et
al.\ 2001; Merritt et al.\ 2001) had reinforced the idea that black holes are
associated with bulges (e.g.\ Dressler \& Richstone 1988; Kormendy \& Gebhardt
2001), bulgeless galaxies with massive black holes have since been detected
(e.g.\ Reines et al.\ 2011; Secrest et al. 2012; Schramm et al.\ 2013; Simmons et al.\ 2013;
Satyapal et al.\ 2014).
Obviously these galaxies do not (yet?) participate in the observed $M_{\rm
  bh}$--$M_{\rm sph,*}$ scaling relation.  As noted in Graham \& Scott (2013),
there are however tens of galaxies known to contain AGN in bulges whose
spheroid magnitudes suggest, based on this near-quadratic $M_{\rm
  bh}$--$M_{\rm sph,*}$ scaling relation, that they harbour intermediate mass
black holes ($10^2 < M_{\rm bh}/M_{\odot} < 10^5$).  It will be interesting to
see a) if this missing population of intermediate-mass black holes exists and b) where they
reside in the $M_{\rm bh}$--$M_{\rm sph}$ diagram.

\subsubsection{Implications} 

Of course the above represents a dramatic revision to the bulge-(black hole)
connection , i.e.\ a completely different relation connecting supermassive
black holes with their host bulges, and as such has wide-spread implications.
For one, the many-merger scenario proposed by Peng (2007), and explored
further by Jahnke \& Macci{\`o} 2011 and Hirschmann et al.\ (2010), to produce
a linear one-to-one scaling via the central limit theorem can be ruled out.
Using a sample of galaxies with a range of initial $M_{\rm bh}/M_{\rm gal,*}$
mass ratios, Peng (2007) noted that after many mergers it would naturally
create an $M_{\rm bh}$--$M_{sph,*}$ relation with a slope of 1.  Although this
concept was independently ruled out by Angl{\'e}s-Alc{\'a}zar et al.\ (2013)
who had emphasized that the number of actual major mergers are not frequent
enough to have established such a linear relation, the quadratic slope of the
$M_{\rm bh}$--$M_{\rm sph}$ relation confirms this ruling.

Some additional implications of the new relation 
%
%
include obvious things like 
(i) black hole mass predictions in other galaxies, 
(ii) estimates of the local black hole mass function 
(e.g.\ Shankar et al.\ 2004,2012; Comastri et al.\ 2015) and mass density based on
local spheroid luminosity functions, and 
(iii) evolutionary studies of the $M_{\rm bh}/M_{\rm sph}$ mass ratio over different
cosmic epochs.  In particular, the local $M_{\rm bh}/M_{\rm sph}$ ratio was thought
to be 0.14\%--0.2\% (e.g., Ho 1999; Kormendy 2001; Marconi \& Hunt 2003;
H\"aring \& Rix 2004).  However Graham (2012a) reported a larger 
value of 0.36\% for the core-S\'ersic galaxies, which was, as noted above, increased that
same year to 0.49\% by Graham \& Scott (2013)\footnote{This announcement appeared
on arxiv.org in mid-November 2012}.  Nearly a year later this higher ratio 
for massive spheroids was again noted in the review by Kormendy \& Ho
(2013) due to its significance. 

Addiionally impacted areas of research include 
(iv) galaxy/black hole formation theories, which extends to
(v) AGN feedback models, 
(vi) predictions for space-based gravitational wave detections,
(vii) connections with nuclear star cluster scaling relations, 
(viii) derivations of past quasar accretion efficiency as a function of mass
(e.g.\ Shankar et al.\ 2009b), 
(ix) searches for the fundamental, rather than secondary, black hole scaling
relation, and 
(x) calibrations matching inactive galaxy samples with 
low-mass AGN data to determine the optimal virial factor for measuring black
hole masses in AGN.  
Given that most of these topics could generate a review in their own right,
only feedback is briefly commented on here. 

A large number of clever theoretical papers have tried to explain the nature
of the $M_{\rm bh}$--$M_{\rm sph}$ relation in terms of feedback from the AGN
(e.g.\ Silk \& Rees 1998; Haehnelt, Natarajan \& Rees 1998; Fabian 1999;
Kauffmann \& Haehnelt 2000; Wilman, Fabian \& Nulsen 2000; Benson et
al.\ 2003; Wyithe \& Loeb 2003; Granato et al. 2004; Di Matteo et al.\ 2005;
Springel et al.\ 2005; Hopkins et al.\ 2005, 2006; Cattaneo et al.\ 2006;
Sijacki et al.\ 2007; Somerville et al.\ 2008; Booth \& Schaye 2009, to
mention just a fraction).  Some papers (but not all those listed here) 
which have claimed success because they 
obtained, through gaseous processes, a linear $M_{\rm bh}$--$M_{\rm sph}$
relation over a wide range of mass, now appear in need of tweaking.  Encouragingly, while
not quite finding a quadratic relation with slope of 2, Hopkins \& Quataert
(2010) report that the black hole growth rate in their models is proportional
to the 1.43 (=1/0.7) power of the star formation rate.

The so-called `quasar' or `cold' mode of black hole growth during gas-rich
processes, as implemented in semi-analytical models, has typically assumed
that the growth occurs via accretion which is linearly proportional to
the inflowing mass of cold gas (which also produces the host spheroid),
modulated by an efficiency which is lower for both unequal mass mergers
(Croton et al.\ 2006) and less massive (more gas-rich) systems with lower
virial velocities (e.g., Kauffmann \& Haehnelt 2000, their eq~2; Croton et
al.\ 2006, their eq.~8; Guo et al.\ 2011, their eq.~36)\footnote{Note: Guo et
  al.\ (2011) excluded the square on the normalised velocity term in their
  eq.~36.}.  Graham \& Scott (2013) therefore presented a new prescription for
the increase in black hole mass, due to gas accretion during wet mergers, such
that the black hole would grow quadratically relative to the host spheroid.
The short duty (on) cycle of quasars ($\sim 10^7$--$10^8$ years) may then
imply that the bulk of a spheroid's stars are also formed rapidly.  Once the
gas is largely gone, and significant galaxy/(black hole) growth is attained
via major dry merger events, the low-accretion model (e.g., Blandford \&
Begelman 1999) presumably results in the so-called `mechanical' or `radio
mode' feedback maintaining the spheroid-(black hole) mass ratio, as is roughly
observed for the core-S\'ersic galaxies.

\begin{greybox}
{ 
\begin{center}
{\bf Updates} 
\end{center}
Measurements of
black hole masses that include the impact of a dark matter halo on the
observed galaxy dynamics have led to 
the upward revision of some $M_{\rm bh}$ estimates.  While this increased the
black hole mass in M87 by a factor of 2 (Gebhardt \& Thomas 2009), the impact
on other galaxies has not only been shown to be less than a factor of 2, but
the 1-sigma uncertainties on the new masses encompass the old values.  That
is, no significant change of mass. For example, the change in mass was a mere
2\% and $-$5\% for NGC~3608 and NGC~4291, just 0.08 dex for NGC~3377 and
NGC~5845, and 0.21 dex for NGC~821 (Schulze \& Gebhardt et al.\ 2011).  Using
these slightly revised masses, plus 4 extra galaxies from the then
newly-published Rusli et al.\ (2013a) paper, and excluding several other
published black hole masses for a plethora of reasons, Kormendy \& Ho (2013)
subsequently reported the 
calibration mid-point of their $M_{\rm bh}$--$M_{\rm sph}$ relation for large
spheroids to be 
$M_{\rm bh}/M_{\rm sph} = $ 0.49\%. 
This agreement with the previously reported mass ratio 
is not particularly surprising given that the masses used by
Kormendy \& Ho (2013) for the large number of galaxies in common with Graham
\& Scott (2013) differed by more than 0.2 dex for just 8 galaxies, and by more
than 0.3 dex for only 5 galaxies.

There are of course two quantities that define the $M_{\rm bh}$--$M_{\rm sph}$
relation, and Savorgnan \& Graham (2016a) have recently completed a thorough
analysis of the spheroid masses for the galaxies listed in Graham \& Scott
(2013) and Rusli et al.\ (2013).  Savorgnan \& Graham (2016a) not only
explain for every galaxy why many published spheroid masses have often
disagreed --- invariably due to inadequate bulge/disc/etc.\ decompositions ---
but they performed the most careful galaxy decompositions to date, effectively
reclassifying many galaxies' morphological type, a process started in Graham
\& Scott (2013, 2015) --- which also included the use of accurate distances to each
galaxy. 
Galaxy reclassification typically occurred when 
a disk or a bar had been over-looked (e.g.\ Graham, Ciambur \& Soria
2016 and Graham et al.\ 2016), or when the contribution from a disk had been
over-estimated in a disky ES type galaxy (as also discussed in Savorgnan \& Graham
2016b and Graham, Ciambur \& Savorgnan 2016).  The new 2016 spheroid masses,
derived from 3.6 $\mu$m images which are not affected by dust obscuration, 
supercede past efforts on many fronts (see Savorgnan \& Graham 2016a for
details).  The revised $M_{\rm bh}$--$M_{\rm sph}$ relation for large
spheroids in early-type galaxies still has a slope consistent with unity,
while the median $M_{\rm bh}/M_{\rm sph}$ ratio has risen to $0.68\pm0.04$\%.
}
\end{greybox}

\begin{greybox}
{
Intriguingly, while the median $M_{\rm bh}/M_{\rm sph}$ ratio has increased,
two points should be made.  As warned by Merritt (2013), the importance of
being able to better resolve the black hole's sphere-of-influence was
illustrated with NGC~1277, whose black hole mass measurement dropped by an
order of magnitude as the spatial resolution increased by an order of
magnitude (van den Bosch et al.\ 2012; Emsellem 2013; Walsh et al.\ 2016;
Graham et al.\ 2016).  
Second, Batcheldor (2010) and Shankar et al.\ (2016) have suggested that the 
sample of galaxies with directly measured black hole masses may reflect the
upper envelope of points in the $M_{\rm bh}$--$M_{\rm sph}$ diagram, because
it is preferentially galaxies with a bigger black hole and thus a bigger
sphere-of-influence that can have their black hole mass measured. 

At the low-mass end of the distribution in the $M_{\rm bh}$--$M_{\rm sph}$
diagram, defined by the bulges of 17 late-type galaxies in Savorgnan et
al.\ (2016), the logarithmic slope of the relation varies from 2 to 3
depending on the type of linear regression used.  This steeper slope (than
observed at the high-mass end) is required for consistency with a wide body of
literature, as we shall see in the coming sections.  We are in the process of
acquiring yet further reliable spheroid masses as there are now (as of
mid-2016) 126 galaxies with directly measured black hole masses, including
close to 50 spiral galaxies.  As discussed at the 2012 IAU 
General Assembly Special Session 3, ``Galaxy Evolution Through Secular
Processes'', Graham (2015b) notes many reasons why pseudobulges cannot be
reliably identified.  Aside from the observation that galaxies can have both a
classical bulge and a pseudobulge --- thus voiding attempts to subsequently bin
galaxies according to whether they have one type or the other --- the fact that a
continuity of morphological criteria exists from high to low bulge masses has
led to considerable confusion regarding the picture of secular versus
non-secular processes (Graham 2014). However, Figure~\ref{Fig2} suggests that
pseudobulges and classical bulges alike are (directly or indirectly) 
broadly aware of their central black hole mass.
} 
\end{greybox}

\subsection{The $L_{\rm sph}$--$\sigma$ relation}

\begin{figure}[t]
\includegraphics[scale=.45, angle=-90]{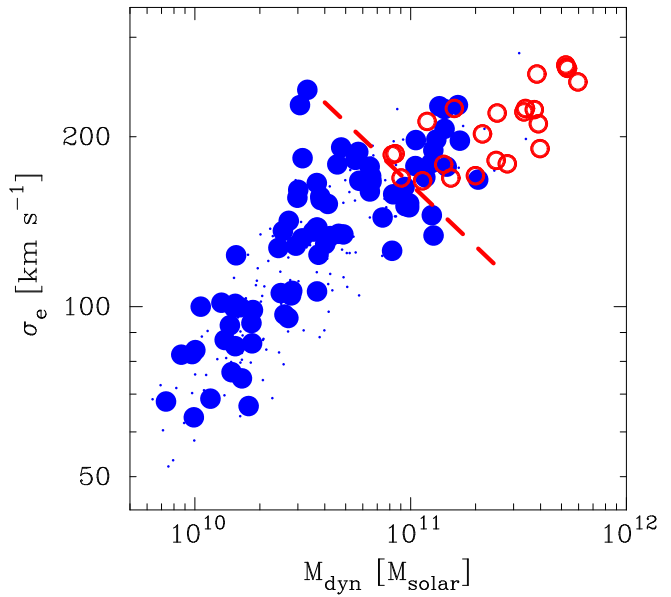}
\caption{Dynamical galaxy mass ($M_{\rm dyn}$) --- equal to twice 
 the Jeans Anisotropic Multi-Gaussian-Expansion mass within the
effective half-light radius $R_{\rm e}$ --- versus the velocity
  dispersion $\sigma_{\rm e}$ within $R_{\rm e}$ for the ATLAS$^{\rm 3D}$ 
early-type galaxies (Cappellari et al.\ 2013, see their Fig.1). Core galaxies
  ($\gamma < 0.3$ according to the Nuker model (Grillmair et al.\ 1994; Lauer
  et al.\ 1995) as used by Krajnovi\'c 
  2013) are shown by the large red circles, while galaxies having steeper inner  
  profiles ($\gamma > 0.5$) are shown by the large blue dots.  Galaxies with
  an unknown inner surface brightness profile slope, or those with 
 $0.3 < \gamma < 0.5$ are shown by the small dots. 
} 
\label{Fig3}
\end{figure}

Around the time that quasars were identified to be at large redshifts,
Minkowski (1962) discovered a correlation between velocity dispersion and
absolute magnitude for early-type galaxies.  He refrained from fitting an
equation to it, noting the need to extend the observations to low absolute
magnitudes.  While Morton \& Chevalier (1973) achieved this, finding a
continuous distribution of velocity dispersions, it was Faber \& Jackson
(1976) who were the first to fit an equation to Minkowski's relation.  For
their sample of 25 galaxies, they reported that $L \propto \sigma^4$, which
has since become known as the Faber-Jackson relation.  A few years later,
exploring the bright end of Minkowski's relation, Schechter (1980) discovered
that $L \propto \sigma^5$, a result confirmed by Malumuth and Kirshner (1981; 
see also von der Linden et al.\ 2007). 
Recent studies have suggested that the exponent may be 5.5 in brightest
cluster galaxies (Liu et al.\ 2008) and as high as 6.5$\pm$1.3 in core galaxies
(Lauer et al.\ 2007).  
Shortly after this, Schechter co-authored Davies et al.\ (1983) in which they
revealed that $L \propto \sigma^2$ for low- and intermediate-luminosity
early-type galaxies.  Many studies have since shown that this result holds
from the lowest luminosity dwarf elliptical galaxies up to $M_B \approx -20$
to $-21$ mag (Held et al.\ 1992; de Rijcke et al.\ 2005; Matkovi\'c \&
Guzm\'an 2005; Balcells et al.\ 2007b; Lauer et al.\ 2007; Chilingarian et al.\ 2008; Forbes
et al.\ 2008; Cody et al.\ 2009; Tortora et al.\ 2009; Kourkchi et
al.\ 2012).  
This explained why past samples of intermediate-to-bright
early-type galaxies had a slope of around 4, or 3 (Tonry 1981), 
and confirmed the observation by Binney (1982) and Farouki et al.\ (1983)
that a single power-law was not appropriate to describe the distribution of
early-type galaxies in the $L$--$\sigma$ diagram. 
Most recently, Davies has again illustrated this 
bend, this time in the $M_{\rm gal}$--$\sigma$ diagram for early-type
galaxies, through co-authorship of Cappellari et al.\ (2013).  Their bent
$M_{\rm gal}$--$\sigma$ diagram is
reproduced in Figure~\ref{Fig3}.  

The bend in Minkowski's relation has been explained by Matkovi\'c \& Guzm\'an
(2005) in terms of S\'ersic galaxies (which have low- and
intermediate-luminosity) following the $L \propto \sigma^2$ relation of Davies
et al.\ (1983) while core-S\'ersic galaxies (which have high-luminosity)
follow the $L \propto \sigma^5$ relation of Schechter (1980).  This continuity
for the low- and intermediate-luminosity S\'ersic galaxies, and the break-away
of bright galaxies with partially depleted cores, is illustrated further in
the $L$--$\mu_0$ and $L$--$n$ distributions seen in Graham \& Guzm\'an (2003,
their figures~9c and 10; see also C\^ot\'e et al.\ 2007, their figure~3e).  As
noted in footnote~\ref{foot_Bekki} of this article, some galaxies may have
experienced a major dry merger event but not display a partially depleted core
--- such as the merger remnants NGC~1316 (Fornax) and NGC~3115 (Schauer et
al.\ 2014; Menezes et al.\ 2014) --- which could explain why some of the
high-mass galaxies in Figure~\ref{Fig3} do not have depleted cores\footnote{It
  will be interesting in the future to careful apply the core-S\'ersic model
  to see how all the points are distributed in terms of galaxies with and
  without partially-depleted cores.}.

The bend in the $M_{\rm gal}$--$\sigma$ diagram, and the $M_{\rm bh}$--$M_{\rm
  sph}$ diagram, is likely to have ties with the flattening that is also
observed at the bright end of the colour magnitude diagram for early-type
galaxies (Tremonti et al.\ 2004; Jim\'enez et al.\ 2011).  Dry merging will
increase the luminosity while preserving the colour (modulo passive evolution)
among the core-S\'ersic elliptical galaxies.  In contrast, the S\'ersic
early-type galaxies display a continuous mass-metallicity relation which
unites the dwarf and ordinary early-type galaxies (e.g.\ Caldwell 1983;
Caldwell \& Bothun 1987).

If the $M_{\rm bh}$--$\sigma$ relation (Section~\ref{Sec_M-sigma}) is roughly 
described by a single power-law, and given that the $L$--$\sigma$ (and $M_{\rm
  gal}$--$\sigma$) relation is notably bent (Figure~\ref{Fig3}), then the $M_{\rm
  bh}$--$L$ relation has to be bent, just as observed and discussed in
Figure~\ref{Fig2} and Section~\ref{SubSec_bend}.

\section{The $M_{\rm bh}$--$\sigma$ relation}
\label{Sec_M-sigma}

While the work on the $M_{\rm bh}$--$L$ relation from Magorrian received
considerable attention, it was the $M_{\rm bh}$--$\sigma$ relation (Ferrarese
\& Merritt 2000; Gebhardt et al.\ 2000) which really sparked off wide-spread
global interest in black hole scaling relations.  The reason may likely have
been because, after having identified and removed galaxies with less secure
black hole mass estimates, the $M_{\rm bh}$--$\sigma$ relation was reported
by both teams to be consistent with having zero intrinsic scatter (see also
Kormendy \& Gebhardt 2001)\footnote{The $M_{\rm bh}$--$L$ relation was
  reported to have more scatter, but this was in part because of poor
  bulge/disc decompositions, and the unrecognised bend in the relation.}.  
That is, after accounting for the measurement errors, all the
scatter was accounted for, suggesting that a new law of physics had been
discovered.  However, the slope of this potential new law was not agreed
upon. Ferrarese \& Merritt (2000) had reported $M_{\rm bh} \propto 
\sigma^{4.8\pm0.5}$, while Gebhardt et al.\ (2000) reported an exponent of
$3.75\pm0.3$.  The former slope agreed with the energy-balancing prediction by
Silk \& Rees (1998, see also Haehnelt, Natarajan \& Rees 1998) that $M_{\rm
  bh} \propto \sigma^5$, while the latter slope agreed with the 
momentum-balancing prediction by Fabian (1999) that $M_{\rm bh} \propto
\sigma^4$.  This discrepancy was to become a major source of controversy and
uncertainty in what has become one of the most famous astronomical relations
of recent years.  As such, some space is dedicated to this issue here. 
In the following subsection, the main reason for the different
slopes is presented, as this continues to be somewhat misunderstood today.

\subsection{Slippery slopes}

Ferrarese \& Merritt (2000) performed a symmetrical linear regression, using
the {\sc bces} routine from Akritas \& Bershady (1996) which allowed for
intrinsic scatter and unique measurement errors on both variables, $M_{\rm
  bh}$ and $\sigma$ (which they took to be 13\% for the velocity dispersion of
external galaxies).  Gebhardt et al.\ (2000), on the other hand, performed a
non-symmetrical ordinary least squares regression by minimising the vertical
offsets (i.e.\ in the $\log M_{\rm bh}$ direction) about their $M_{\rm
  bh}$--$\sigma$ relation.  This approach effectively assumed that the
uncertainty on the velocity dispersion was zero and that the black hole masses
all had the same uncertainty.

Merritt \& Ferrarese (2001a) addressed the issue of the differing slopes, using
four different types of linear regression, two which treated the ($M_{\rm
  bh}$, $\sigma$) data symmetrically and two which did not.  They revealed how
the slope of the $M_{\rm bh}$--$\sigma$ relation increased as one assigned an
increasing uncertainty to the velocity dispersion and presented a best fit
slope of 4.72$\pm$0.36 for their expanded sample.

Tremaine et al.\ (2002) also looked at this issue of different slopes and
noted that under certain conditions\footnote{The slope can be biased if (i)
  the uncertainty on the $x$ values is large compared to the range of $x$
  values, or (ii) the sizes of all the $x$ and $y$ uncertainties are not
  roughly comparable to each other.} the minimisation routine from Akritas \&
Bershady, which was used by Ferrarese \& Merritt (2000), can be biased.  As
noted above, Merritt \& Ferrarese (2001a) had additionally used a second
symmetrical regression routine, referred to as the ``Orthogonal distance
regression'' which had been implemented by Press et al.\ (1992, their Section~15.3) as {\sc
  FITEXY}.  It was such that the following quantity was minimised during the
task of fitting the line $y=a+bx$ 
\begin{equation}
\chi^2 = \sum_{i=1}^N \frac{[y_i-(a + bx_i)]^2}
    { {\delta y_i}^2 + b^2{\delta x_i}^2 }, 
\label{Eq_One}
\end{equation}
where $N$ data pairs of $y$ and $x$ values are available in one's sample, 
and they have measurement errors $\delta y$ and $\delta x$, respectively. 
Merritt \& Ferrarese (2001a) pointed out that Feigelson \& Babu (1992) had 
already noted that this routine is fine unless the distribution to be fit contains
intrinsic scatter, i.e.\ real departures of the data from the optimal line which are
not due to measurement errors.  At that time, the $M_{\rm bh}$--$\sigma$
relation was thought to contain no intrinsic scatter, or was at least
consistent with having no intrinsic scatter. 

Tremaine et al.\ (2002) subsequently developed their own modified version of 
{\sc FITEXY}.  I t was such that it minimised the quantity 
\begin{equation}
\chi^2 = \sum_{i=1}^N \frac{[y_i-(a + bx_i)]^2}
    { {\delta y_i}^2 + b^2{\delta x_i}^2 + \epsilon_y^2 }, 
\label{Eq_Two}
\end{equation}
where the intrinsic scatter $\epsilon_y$ is solved for by repeating the fit
until $\chi^2/(N-2)$ equals 1.  Although Tremaine et al.\ (2002) claimed this
expression still gave a symmetrical treatment of the data, it did not.  By
trying to allow for intrinsic scatter, they
had inadvertently converted a symmetrical expression into a non-symmetrical
expression by minimising the offsets under the assumption that all of the
intrinsic scatter lay in the $y$-direction.
They reported a slope of $4.02\pm0.32$ for their $M_{\rm bh}$--$\sigma$
relation using the smaller uncertainty 
of 5\% (compare 13\%) for the velocity dispersions of the external galaxies. 

Here we look at this a little more carefully, as it continues to cause 
confusion more than a decade later. 
If one was to minimise the offsets in the $x$-direction, 
about the line $y=a+bx$, or equivalently $x=(y-a)/b$, 
the expression would be
\begin{eqnarray}
\chi^2 & = & \sum_{i=1}^N \frac{[x_i-\frac{(y_i - a)}{b}]^2}
    { {\delta y_i}^2/b^2 + {\delta x_i}^2 + \epsilon_x^2 },  \nonumber \\
      &   & \sum_{i=1}^N \frac{[-y_i+(a + bx_i)]^2}
    { {\delta y_i}^2 + b^2{\delta x_i}^2 + b^2\epsilon_x^2 }, 
\label{Eq_Three}
\end{eqnarray}
where $\epsilon_x$ is the intrinsic scatter, but this time implicitly assumed
to reside in the $x$-direction.  The difference between equations~\ref{Eq_Two}
and \ref{Eq_Three} is the final term in the denominator, which has that
$\epsilon_y = b\epsilon_x$.  Given this (not surprising) dependence on the
slope between $\epsilon_y$ and $\epsilon_x$, the solution reached by solving
for $\chi^2/(N-2)=1$ in equations~\ref{Eq_Two} and \ref{Eq_Three} has a
different value of $b$, i.e.\ a different slope.  To obtain a symmetrical
regression therefore requires an average of these two regressions as discussed
in Novak et al.\ (2006)\footnote{An easy way to 
  check if one has performed a symmetrical regression is to swap their $x$ and
  $y$ data around and re-feed this into their regression routine.}, which are
sometimes referred to as the forward and the inverse regression.

Performing a non-symmetrical linear regression analysis and minimising the
offsets in just the $\log M_{\rm bh}$ direction is preferred if one wishes to
obtain a relation useful for predicting black hole masses in other galaxies,
simply because this relation has the smallest offsets in the $\log M_{\rm bh}$
direction (see Feigelson \& Babu 1992; Andreon \& Hurn 2012).  If, on the other hand, one is
interested in the underlying / fundamental relation connecting $M_{\rm bh}$
and $\sigma$, then one should perform a symmetrical regression.  This is
discussed by Novak et al.\ (2006) in terms of the Observer's Question and the
Theorist's Question.

Analysing the same data\footnote{The black hole mass for NGC~821 was updated,
  but this had almost no impact.}  from Tremaine et al.\ (2002), and assigning
a 5\% uncertainty to the velocity dispersion of each galaxy (including the
Milky Way), Novak et al.\ (2006) reported a slope of 4.10$\pm$0.30 using
Eq.~\ref{Eq_Two} and 4.59$\pm$0.34 using Eq.~\ref{Eq_Three}.  Had they used an
uncertainty of 13\%, they would have reported slopes of 4.39 and 4.59, giving
an average value slope of 4.49 that was consistent with Merritt \& Ferrarese
(2001a) who reported an optimal slope of 4.72$\pm$0.36.

To make a point about the ongoing concerns regarding different minimisation
routines, and in particular to show that the symmetrical bisector regression
routine from Akritas \& Bershady was not producing a biased fit in regard to
the ($M_{\rm bh}, \sigma$) data, Graham \& Li (2009) used three symmetrical
regression routines, one from Akritas \& Bershady (1996), the expression from
Tremaine et al.\ (2002) operating in both forward and inverse mode, and an IDL
routine from Kelly (2007) based on a Bayesian estimator.  All were shown to
give very similar results when the same uncertainty on the velocity dispersion
was consistently used, a test that was recently confirmed in Park et
al.\ (2012) who additionally used a fourth (maximum likelihood) estimator.

\subsection{substructure and escalating slopes}

In 2007 Graham noticed that all of the barred galaxies in the $M_{\rm
  bh}$--$\sigma$ diagram were offset, to either lower black hole masses and/or
higher velocity dispersions, relative to the best-fitting line defined by the
non-barred galaxies, and that excluding the barred galaxies resulted in a
reduced scatter about the $M_{\rm bh}$--$\sigma$ relation (Graham 2007a).  At
the same time, Hu (2008) had compiled a larger sample and shown the same
apparent substructure within the $M_{\rm bh}$--$\sigma$ diagram.  Hu
considered all of his offset galaxies to contain `pseudobulges', built from
the secular evolution of their surrounding disc and containing relatively under-developed
black holes.  They were also all barred galaxies.  Graham (2008a) similarly
considered the offset galaxies to have undermassive black holes, due to
secular evolution over-developing the bulge, or to have elevated velocity
dispersions due to the dynamics of the bar.  The choice appears answered
because Hartmann et al.\ (2014) have shown that bars are indeed capable of
increasing the velocity dispersion in galaxies, and by exactly the average
offset observed in the $M_{\rm bh}$--$\sigma$ diagram (see also Debattista et
al.\ 2013 and Monari et al.\ 2014).  Furthermore, Figure~\ref{Fig2} shows that
pseudobulges and classical bulges (and clump bulges) follow the same broad
distribution in the $M_{\rm bh}$--$M_{\rm sph}$ diagram; at low spheroid
masses they both reside systematically below the near-linear relation defined
by the massive core-S\'ersic spheroids.  There is not yet evidence that
pseudobulges contain smaller black hole masses than classical bulges of the
same mass, although more data would be welcome. In particular, removing the
contribution of the bar\footnote{Graham et al.\ 2011, their figure~7, offer a
  first order approximation for this.}, and the rotational contribution\footnote{See Kang et
  al.\ 2013, their figure~9, and Pota et al.\ 2013 in regard to the velocity
  dispersion of a globular cluster system.}, from the observed central velocity 
dispersions of the spheroids would be helpful.  It may also make more sense to use 
the quantity $\sqrt{3\sigma^2_{\rm sph} +v^2_{\rm sph,rot}}$ (Busarello et
  al.\ 1992). 
Although, much of this may be moot in regard to 
pseudobulges due to the difficult task of actually identifying them, as
discussed in the following subsection.  

One thing that was clear from Hu (2008) and Graham (2008b) was that the
growing sample size had generated an increased scatter about the $M_{\rm
  bh}$--$\sigma$ relation\footnote{Potentially, this may in part be due to the
  inclusion of less accurate black hole mass measurements with under-estimated
  error bars (see Merritt 2013).}, and the intrinsic scatter no longer
appeared consistent with zero, a result shown further by G{\"u}ltekin et al.\ (2009).
The $M_{\rm bh}$--$\sigma$ diagram was therefore falling from grace, and it also
now presented quite a contrast to early claims which had reported that
classical bulges and pseudobulges follow the same black hole scaling relations
(e.g.\ Kormendy 2001; Kormendy \& Gebhardt 2001).  In Kormendy et al.\ (2011)
the offset nature of the pseudobulges was acknowledged, and it was now claimed that
black hole masses do not correlate with the properties of
pseudobulges. However, the
range in absolute magnitude of the pseudobulges was restricted to just
2 mag, making it challenging to identify if there is a relation present.
With a fuller data set, Figure~\ref{Fig2} reveals that all bulge types appear
to follow an $M_{\rm bh}$--$M_{\rm sph}$ relation. 

With a sample size of 72 galaxies, McConnell \& Ma (2013) used the
non-symmetrical, modified {\sc FITEXY} routine, as coded by Williams et
al.\ (2010) in {\sc MPFITEXY}.  They reported a slope of 5.64$\pm$0.32 for
their optimal $M_{\rm bh}$--$\sigma$ relation (their figure~1, which included
the alleged over-massive black hole in NGC~1277 from van den Bosch et al.\ 2012
which has since been rescinded). 
If they had of additionally used the inverse of this regression, in which the
unknown intrinsic scatter is assigned to the $\log\, \sigma$ direction, they
would have obtained a slope of 6.64, and thus an average slope of 6.14.  This
is steeper than previously reported, and is in part due to their inclusion of
the offset barred galaxies at low masses.  While McConnell \& Ma (2013) do
report that their 19 late-type galaxies (with both classical bulges and
pseudobulges) have an $M_{\rm bh}$--$\sigma$ relation with a zero point
(i.e.\ the term `$a$' in $y=a+bx$) that is 0.29 dex lower than for their 53
early-type galaxies (8.36 vs 8.07), i.e.\ offset by a factor of 2, they did
not perform a fit to the barred and non-barred galaxies. Given that the
early-type galaxies dominate at the high-mass end of the diagram, and the
late-type galaxies at the low-mass end, they combine to produce the steeper
relation with a slope of $\approx$6.

\begin{figure}[t]
\includegraphics[scale=.47,angle=270]{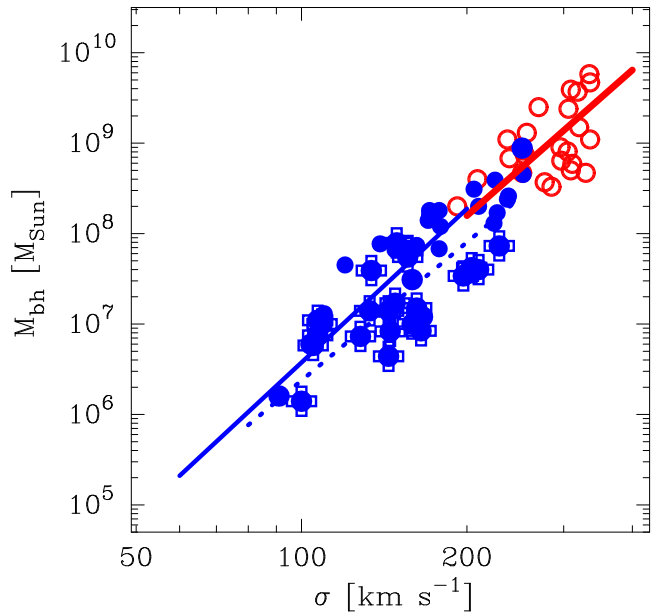}
\caption{$M_{\rm bh}$--$\sigma$ diagram taken from Graham \& Scott (2013).
Red circles represent core-S\'ersic galaxies; blue dots represent 
  S\'ersic galaxies. The crosses designate barred galaxies, which tend to be
  offset to higher velocity dispersions. The three lines are linear
  regressions, in which the barred S\'ersic galaxies and the non-barred
  S\'ersic galaxies have been fit separately from the core-S\'ersic galaxies
  (which are not barred). 
} 
\label{Fig4}
\end{figure}

Graham et al.\ (2011) highlighted a potential sample selection bias such that
the need to resolve (or nearly resolve) the sphere-of-influence of the black
holes may be resulting in an artificial floor to the distribution of points in
the $M_{\rm bh}$--$\sigma$ diagram.  As such, they additionally used a
non-symmetrical regression, but one which minimised the offsets in the
horizontal direction, i.e.\ they performed the `inverse' regression as this
should provide the least biased fit (see Lynden-Bell et al.\ 1988).  Adding
eight black hole masses to the compilation of 64 data pairs in Graham et
al.\ (2011), Graham \& Scott (2013) reported a slope of 6.08$\pm$0.31 using
their preferred inverse regression on their sample of 72 galaxies (see
Figure~\ref{Fig4}).  For the 51 non-barred galaxies, their optimal slope using
the inverse regression was 5.53$\pm$0.34.  While this is at first glance in
agreement with the preferred value of 5.64$\pm$0.32 reported by McConnell \&
Ma 2013, it should be realised that it is a coincidence as different things
have been measured: a forward regression for all galaxy types versus an
inverse regression for non-barred galaxies.

Using updated and expanded data for 57 non-barred galaxies, taken from the
sample of 89 galaxies in Savorgnan \& Graham (2015), the forward, inverse and
average regression give a slope of 5.10, 6.48 and 5.79.  Folding in the offset
barred galaxies results in steeper slopes still, as seen with the McConnell \&
Ma (2013) data.  The increase to the slope over the past few years (see also
Sabra et al.\ 2015 who report a slope of 4.60 using 89 galaxies and the
`forward' linear regression) has largely 
come from increased black hole masses, and new data, at the high mass end.
McConnell \& Ma (2013) additionally note that the flux-weighted velocity
dispersion within one effective radius can be as much as 10--15\% lower in
their massive galaxies when excluding data within the black hole's
sphere-of-influence. This follows Graham et al.\ (2011) who noted that the
velocity dispersion for M32's spheroid should be reduced from $\sim$75 km
s$^{-1}$ to $\sim$55 km s$^{-1}$ (Tonry 1987) for exactly this reason.
Increases to black hole masses have also come from efforts to account for dark
matter halos, resulting in an average increase of $\sim$20\% (Schulze \&
Gebhardt 2011; Rusli et al.\ 2103a), but as high as a factor of 2 in the case
of M87 (Gebhardt \& Thomas 2009). Incorporating a dark matter halo is 
akin to relaxing the past assumption/simplification that the stellar
mass-to-light ratio is constant with radius\footnote{This raises another issue
  which is yet to be properly addressed in the literature: not only do many
  spheroids have radial stellar population gradients, but most S\'ersic
  galaxies have nuclear star clusters in addition to massive black holes, and
  the assumption of a single stellar mass-to-light ratio when modelling the
  data to derive a black hole mass is therefore not appropriate.}. 

This new, slightly steeper, $M_{\rm bh}$--$\sigma$ relation for the non-barred
galaxies suggests that if $L_{\rm sph} \propto \sigma^6$ (Lauer et al.\ 2007) 
for the core-S\'ersic galaxies, then one can expect to recover $M_{\rm bh}
\propto L_{\rm sph}$ for the core-S\'ersic galaxies.  If $L_{\rm
  sph} \propto \sigma^5$ (e.g.\ Schechter 1980) then one can expect to find $M_{\rm bh} \propto
L^{6/5}_{\rm sph}$, suggestive of a second order effect on the picture of dry
mergers maintaining a constant $M_{\rm bh}/L_{\rm sph}$ and $M_{\rm bh}/M_{\rm sph}$
ratio. Resolution to this minor query may simply require consistency with the
regression analyses, or perhaps a careful bulge/disc separation of the
galaxies involved (e.g.\ Laurikainen et al.\ 2005, 2011; Balcells et
al.\ 2007a,b; Gadotti 2008; L\"asker et al.\ 2014a), because core-S\'ersic
galaxies can contain a fast-rotating disc (e.g.\ Dullo \& Graham 2013;
Krajnovi{\'c} et al.\ 2013).


\subsubsection {Pseudobulges \label{Sec_pseudo}}

Pseudobulges are particularly hard to identify, for the multitude of reasons
presented in Graham (2013, 2014).  Furthermore, many galaxies contain {\it
  both} a disc-like `pseudobulge' and a classical bulge (e.g.\ Erwin et
al.\ 2003, 2014; Athanassoula 2005; Gadotti 2009; 
MacArthur, Gonz\'alez \& Courteau 2009; dos Anjos \& da Silva
2013; Seidel et al.\ 2014), including the Milky Way it seems
(e.g.\ D{\'e}k{\'a}ny et al.\ 2013; Kunder et 
al.\ 2016; see also Saha 2015).  In addition, some may have formed from the
(secular) inward migration and (classical) merging of stellar clumps
(e.g.\ Noguchi 1999; Bournaud et al.\ 2007; Inoue \& Saitoh 2012, and
references therein).  All of this makes the task of labelling galaxies as
either containing a pseudobulge or a classical bulge highly problematic and
untenable.  In the $M_{\rm bh}$--$\sigma$ analysis by Graham et al.\ (2011)
and Graham \& Scott (2013), they avoided the issue of pseudobulges and
separated galaxies based on the presence (or not) of a bar and revealed that
the masses of black holes in barred galaxies correlate with the velocity
dispersion, despite their heightened dynamics.  Given that the majority of
S\'ersic spheroids (i.e.\ those without partially depleted cores) also follow
the near-quadratic $M_{\rm bh}$--$L$ relation, it appears that the masses of
black holes in pseudobulges correlate with at least one property of their host
bulge, and unless pseudobulges are restricted to have a narrow range of
velocity dispersion, then their black hole masses also correlate with velocity
dispersion (or at least define an upper envelope in the $M_{\rm bh}$--$\sigma$
diagram).

A few of the (often not properly recognised) difficulties with identifying
pseudobulges are noted here, in case it is helpful to some readers.  
From a kinematical perspective, just as with the formation of rotating
elliptical galaxies via mergers, mergers can also create bulges which rotate
(e.g.\ Bekki 2010; Keselman \& Nusser 2012) and bars can spin-up classical
bulges (e.g.\ Saha et al.\ 2012, 2016), 
and the smaller the bulges are the easier it is.  Rotation
is therefore not a definitive signature of a pseudobulge.  
In spiral galaxies, the observable presence of the disc's
inner spiral arms, which cohabit the inner region of the galaxy where the
bulge also resides, are of course easier to detect in fainter bulges (which
are those that have smaller S\'ersic indices) due to
the greater bulge/arm contrast.  However the detection and presence of these underlying
features does not necessitate the presence of a pseudobulge (e.g.\ Eliche-Moral et
al.\ 2011; dos Anjos \& da Silva 2013).

From a selection of hundreds of disc galaxies imaged in the $K$-band, Graham
\& Worley (2008) observe no bimodality in the bulge S\'ersic indices,
questioning the suitability of a divide at a S\'ersic index of $n=2$ which has
frequently been used in the recent literature.  This divide is roughly halfway
between $n=1$ (which describes the light-profiles of flattened rotating discs)
and $n=4$ (which was in the past thought to describe the majority of
elliptical galaxies and large bulges).  While pseudobulges are expected to
have S\'ersic indices $n \approx 1$ --- having formed from their surrounding
exponential disc (e.g.\ Bardeen 1975; Hohl 1975; Combes \& Sanders 1981;
Combes et al.\ 1990; Pfenniger \& Friedli 1991) --- the problem is that
mergers do not only produce $R^{1/4}$-like light profiles. Mergers can also
create bulges with $n < 2$ (e.g.\ Eliche-Moral et al.\ 2011; Scannapieco et
al.\ 2011; Querejeta et al.\ 2015), just as low-luminosity elliptical galaxies (not built from the
secular evolution of a disc) are well known to have $n<2$ and even $<1$
(e.g.\ Davies et al.\ 1988; Young \& Currie 1994)\footnote{The occurrence of
  large-scale, rotating stellar discs and kinematical substructure in
  early-type galaxies on either side of the alleged divide at $M_B = -18$ mag
  ($n\approx 2$) further reveals the continuity of dwarf and ordinary
  early-type galaxies (e.g., Emsellem et al.\ 2007; Krajnovi\'c et al.\ 2008;
  Scott et al.\ 2014; Toloba et al.\ 2014).}.

Prior to the realisation that the S\'ersic index changes monotonically with
spheroid luminosity and size (e.g.\ Caon et al.\ 1993; Andredakis et
al.\ 1995) --- referred to as structural nonhomology --- the curved but
continuous scaling relations involving the 'effective' half-light radii and
`effective' surface brightness (which have a maximum curvature around $n=2$)
had suggested that spheroids with $n<2$ may be a distinct species rather than
the low mass extension of spheroids with $n>2$ (see Graham 2013).  However we
now know that this was a red-herring, and that all relations involving the
`effective' parameters are curved (e.g.\ Graham \& Guzm\'an 2003; Gavazzi et
al.\ 2005; Ferrarese et al.\ 2006a; C\^ot\'e et al.\ 2006, 2007).  As such,
the Kormendy (1977) relation cannot be used to separate dwarf early-type
galaxies from ordinary early-type galaxies, nor to separate pseudobulges from
classical bulges, because at low-luminosities both types of bulge (classical
and pseudo) depart from this relation, which is the tangent to the bright arm
of the curved $\mu_{\rm e}$--$R_{\rm e}$ distribution.

\section{The $M_{\rm bh}$--$n$ relation}
\label{Sec_n}

As noted in Graham et al.\ (2001), it may not be the total amount of mass in a
spheroid, but rather how that mass is distributed, when it comes to the
connection with the central supermassive black hole.  Similarly, the velocity
dispersion is but a tracer of the underlying mass distribution, and as such
it can not be the fundamental parameter driving the black hole mass
scaling relations.  

\begin{figure*}
\plottwo{Fig5a.ps}{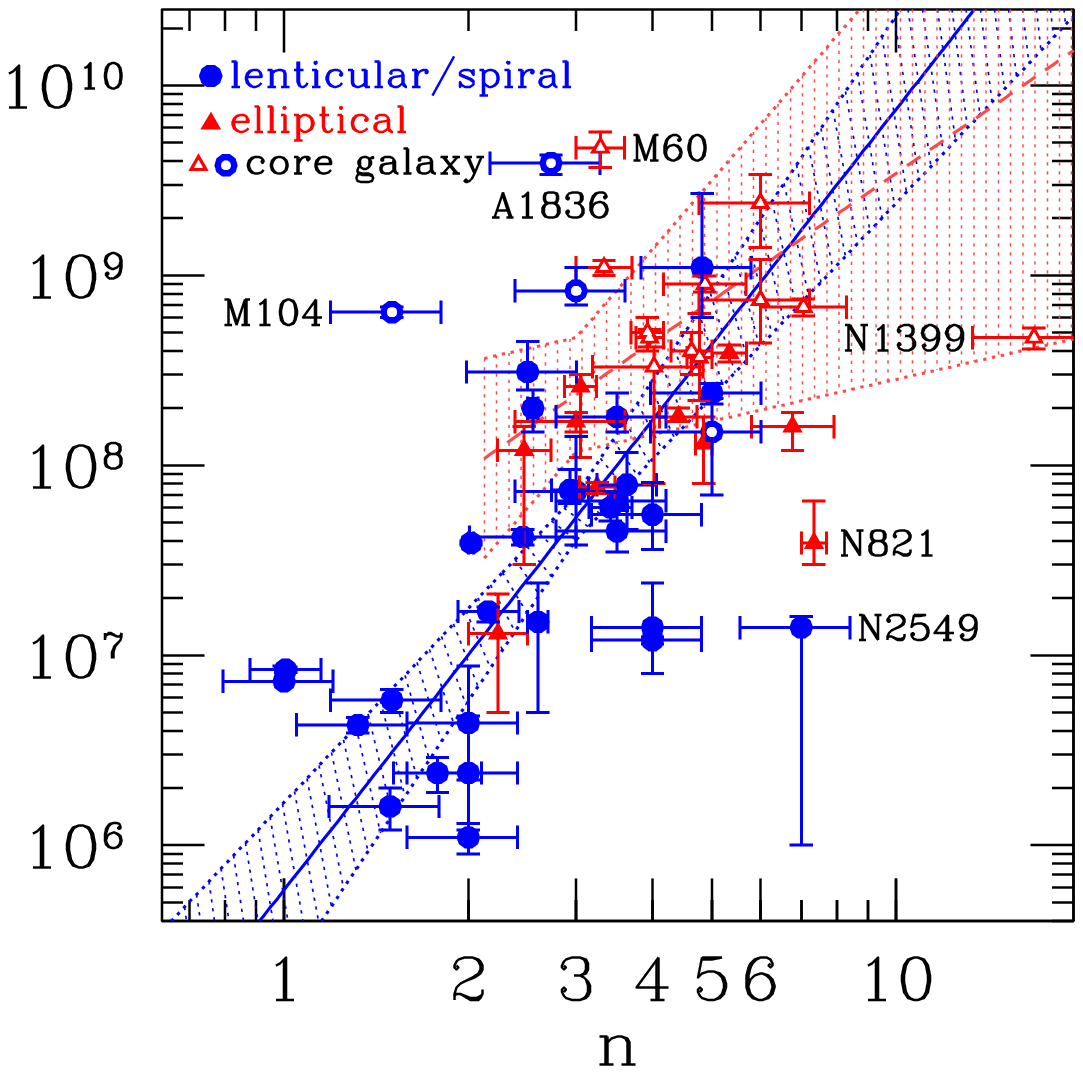}
\caption{Left panel: $M_{\rm bh}$--$n$ diagram taken from Graham \& Driver
  (2007).  The core-S\'ersic spheroids are shown here by the red circles,
  while the S\'ersic spheroids are shown by the blue dots. 
The lone S\'ersic spheroid at the high-mass end is the S0 galaxy NGC~3115,
identified to not have a core by Ravindranath et al.\ (2001). 
Right panel: $M_{\rm bh}$--$n$ diagram from Savorgnan et al.\ (2013).
Rather than a single log-quadratic relation, two log-linear relations are
shown here, one for the S\'ersic spheroids and one for the core-S\'ersic
spheroids. 
}
\label{Fig5}
\end{figure*}

Intriguingly, what Graham et al.\ (2001) revealed is that the central radial
concentration of light, within the inner effective half light radii of
spheroids, correlates strongly with the black hole mass.  The concentration
index which they used, taken from Trujillo et al.\ (2001), is monotonically
related with the S\'ersic index $n$, and thus an $M_{\rm bh}$--$n$ relation
also exists, as shown in Graham et al.\ (2003a).  With an expanded data set,
Graham \& Driver (2007) revealed that this relation is no longer well
described by a single log-linear power-law, and that a log-quadratic relation
performs noticeably better (see Figure~\ref{Fig5}a).  Given the log-linear
$L$--$n$ relation observed for both elliptical galaxies (e.g.\ Young \& Currie
1994; Jerjen \& Binggeli 1997; Graham \& Guzm\'an 2003; Ferrarese et
al.\ 2006a)
and the bulges of disc galaxies (e.g.\ Andredakis et al.\ 1995;
Graham \& Worley 2008, and references therein), and the bent $M_{\rm
  bh}$--$L_{\rm sph}$ relation (Section~\ref{Sec_M-M}), the $M_{\rm bh}$--$n$
relation must be bent, such that galaxies which have experienced major,
relatively dry, merger events are responsible for the flattening which is seen
in Figure~\ref{Fig5} at high masses.

The existence of the $M_{\rm bh}$--$L_{\rm sph}$ relation, coupled with 
existence of the $L_{\rm sph}$--$n$ relation,  
necessitates the existence of the $M_{\rm bh}$--$n$ relation.  Although, 
as illustrated by Savorgnan et al.\ (2013), there is a need for care when
measuring S\'ersic indices, and studies which fail to recover the 
$M_{\rm bh}$--$n$ relation for the sample of galaxies with directly measured
black hole masses 
may be dominated by poorly measured S\'ersic indices, and in turn erroneous bulge
magnitudes which depend on an accurate S\'ersic index. 
Within the literature, measurements for 
individual galaxies have varied dramatically (e.g.\ Graham \& Driver 2007;
Laurikainen et al.\ 2010; Sani et al.\ 2011; Vika et al.\ 2012; Beifiori et
al.\ 2012; Rusli et al.\ 2013a; L\'asker et al.\ 2014a).  Shown in
Figure~\ref{Fig5}b are the average values, after the rejection of extreme
outliers, plotted against black hole mass.  Savorgnan et al.\ (2013) divided
the sample into S\'ersic and core-S\'ersic spheroids, and fit separate linear
regressions for each sub-population.

Savorgnan (2016, in prep.) will present an investigation based on a careful 
multi-component analyses (of the 72 galaxies used by Graham \& Scott 2013)
which reconciled the differences between past attempts to measure the S\'ersic
index.  For example, sometimes these discrepancies arise because a lenticular disc galaxy
may have been modelled with either a single S\'ersic component or more
correctly as the sum of a S\'ersic-bulge plus an exponential disc by a
different author.  Other times the presence of an unaccounted for nuclear
disc, or a partially depleted core, has biased the luminosity-weighted fits in
some studies.  
Despite the need for care when measuring the S\'ersic index, the
advantage is that one only requires uncalibrated photometric images. 

Readers interested in the development of fitting bulge light profiles since de
Vaucouleurs (1959) first noted departures from his $R^{1/4}$ model, may
appreciate the references in section 4.1 of Graham (2013).  Andredakis et
al.\ (1995) were the first to model the bulges of disc galaxies with
S\'ersic's (1963) light profile model, following its application to
elliptical galaxies by Davies et al.\ (1988) and Caon et al.\ (1993), and the
earlier advocation of its use by Capaccioli (1985, 1987).  Some of the
difficulty with, and the impact of getting, the S\'ersic index correct is
illustrated by Gadotti \& S{\'a}nchez-Janssen (2012) in the case of the
Sombrero galaxy.

\section{The $M_{\rm bh}-\mu_0$ diagram}
\label{Sec_mu0}

It is not unreasonable to expect that the growth of massive black holes may be
related to the growth, and subsequent space density, of stars in its immediate
vicinity. Gas processes have contributed to the development of both, and the
black hole mass may be more connected with the local stellar density than the
total stellar mass of the host spheroid.  While the de-projected stellar
density, $\rho_0$ is ideally the quantity we would like to have (e.g.\ Merritt
2006b, his figure~5), and this can be derived under certain assumptions
(e.g.\ Terzi\'c \& Graham 2005, their Eq.~4), it is of course the projected
surface brightness that is observed.

Binggeli, Sandage \& Tarenghi (1984) and Sandage \& Binggeli (1984) provide a
nice historical account of the detection of dwarf galaxies, and wrote that it
was established that ``the dwarf elliptical galaxies form a continuum in
luminosity with the brighter E systems''.  Caldwell (1983; his Figure 6) and
Bothun et al.\ (1986, their figure~7) revealed this continuum was such that
fainter than $M_B \approx -20.5$ mag, there is a log-linear relation between
the luminosity and the central surface brightness, $\mu_0$.  In addition to
this, Binggeli et al.\ (1984, their figure~11) and Binggeli \& Cameron (1991,
their figures~9 and 18) found that, when using the inward extrapolation of
King models, this $L$--$\mu_0$ relation extends from $-12 > M_B > -23$ mag.
This was further highlighted by Jerjen \& Binggeli (1997) and Graham \&
Guzm\'an (2003) when using the inward extrapolation of the S\'ersic model;
extrapolated over partially depleted cores in the case of the brightest
spheroids whose cores have been eroded away by coalescing supermassive black
holes.

\begin{figure}[b]
\includegraphics[scale=0.82,angle=270]{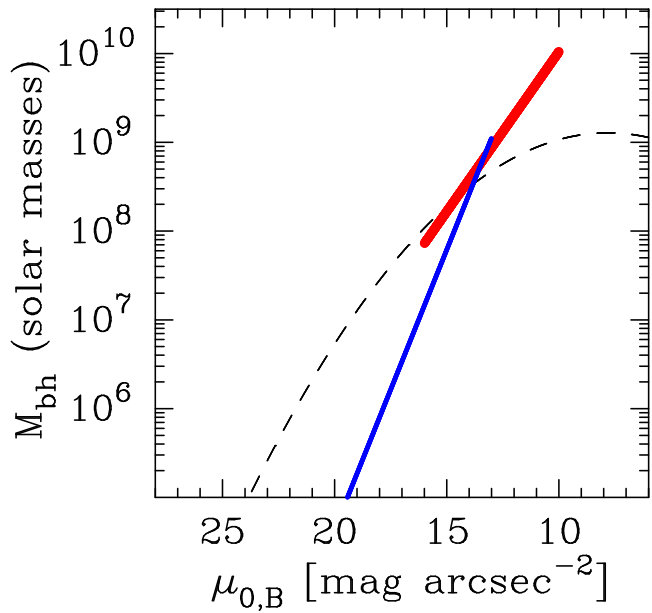}
\caption{Predictions for the $M_{\rm bh}$--$\mu_0$ diagram.
The dashed curve is from Graham \& Driver (2007), while the thin blue and
thick red lines
show equations~\ref{Eq_Ser_mu} and \ref{Eq_cS_mu} for the S\'ersic and
core-S\'ersic spheroids, respectively. Clearly the uncertainty on these lines
is still quite large, given that the solid lines do not trace the dashed
curve, but a bend is nonetheless expected.} 
\label{Fig6}
\end{figure}

Given this log-linear $L$--$\mu_0$ relation, and the bent $M_{\rm
  bh}$--$L_{\rm sph}$ relation (Section~\ref{Sec_M-M}), there must be a bent
$M_{\rm bh}$--$\mu_0$ relation.  It should again be emphasized that this
particular value of $\mu_0$ refers to the extrapolated / expected value prior
to core depletion.  Given the difficulties in routinely obtaining robust
S\'ersic indices for the spheroids with black hole masses
(Section~\ref{Sec_n}), it is perhaps not surprising that this diagram is yet to
be published.  Although it may be the fundamental parameter linking black
holes with their bulges, to date there is only a prediction by Graham \&
Driver (2007) for its form.  This was derived by coupling the log-quadratic
$M_{\rm bh}$--$n$ relation from Graham \& Driver with the log-linear
$n$--$\mu_0$ relation from Graham \& Guzm\'an (2003), and is reproduced here
in Figure~\ref{Fig6}.

Given our current understanding, it makes more sense to construct the $M_{\rm
  bh}$--$\mu_0$ relation using the log-linear $M_{\rm bh}$--$L$ relations for
the S\'ersic and core-S\'ersic spheroids given in Graham \& Scott (2013, their
table~3) together with the log-linear $L$--$\mu_0$ relation given in Graham \&
Guzm\'an (2003, their figure~9c).  Because the latter was derived in the
$B$-band, we use the $B$-band $M_{\rm bh}$--$L$ relation from Graham \& Scott
(2013).  For the S\'ersic galaxies, this gives the relation 
\begin{equation}
\log (M_{\rm bh}/M_{\odot}) = 17.24 - 0.63\mu_0, 
\label{Eq_Ser_mu}
\end{equation}
and for the core-S\'ersic galaxies one has the relation
\begin{equation}
\log (M_{\rm bh}/M_{\odot}) = 13.62 - 0.36\mu_0. 
\label{Eq_cS_mu}
\end{equation}
These predictions are shown in Figure~\ref{Fig6}. 
From the multi-component modelling by Savorgnan \& Graham (2016a) 
of galaxies with directly measured black hole masses, it will be possible
to populate this diagram and (under certain assumptions) its deprojected
cousin.

\section{Depleted galaxy cores and the $M_{\rm bh}$--$M_{\rm def}$ relation}
\label{Sec_core}

As noted previously, the merger of two galaxies without substantial gas,
referred to as a dry merger, will result in the supermassive black holes from
the progenitor galaxies sinking to the bottom of the newly wed galaxy by
transferring much of their orbital angular momentum to the stars near the new
galaxy's core (Begelman, Blandford \& Rees 1980; Ebisuzaki et al.\ 1991).
Such collisional construction of galaxies results in an evacuated `loss cone'
showing up as a partially depleted core\footnote{See Dullo \& Graham (2013,
  their Section~6.1) for a discussion of alternative concepts for core
  depletion.} 
in the images of nearby galaxies
(e.g.\ King \& Minkowski 1966, 1972; Kormendy 1982; Lauer 1983).  Typical core
sizes, as quantified by the break radius $R_b$ of the core-S\'ersic model, are
tens to a few hundred parsec (e.g.\ Trujillo et al.\ 2004; Ferrarese et
al.\ 2006a; C\^ot\'e et al.\ 2007; Hyde et al.\ 2008; Richings et al.\ 2011;
Rusli et al.\ 2013b; Dullo \& Graham 2013, 2014; Bonfini 2014), and roughly a
factor of 2 smaller than Nuker model break radii (Lauer et al.\ 1995).
Whether or not coalescence of the black holes has already occurred in these
galaxies with partially depleted cores is not clear, although see Khan et
al.\ (2011, 2013, and references therein) in regard to the `final parsec problem'. 

\begin{figure}
\includegraphics[scale=0.195, angle=0]{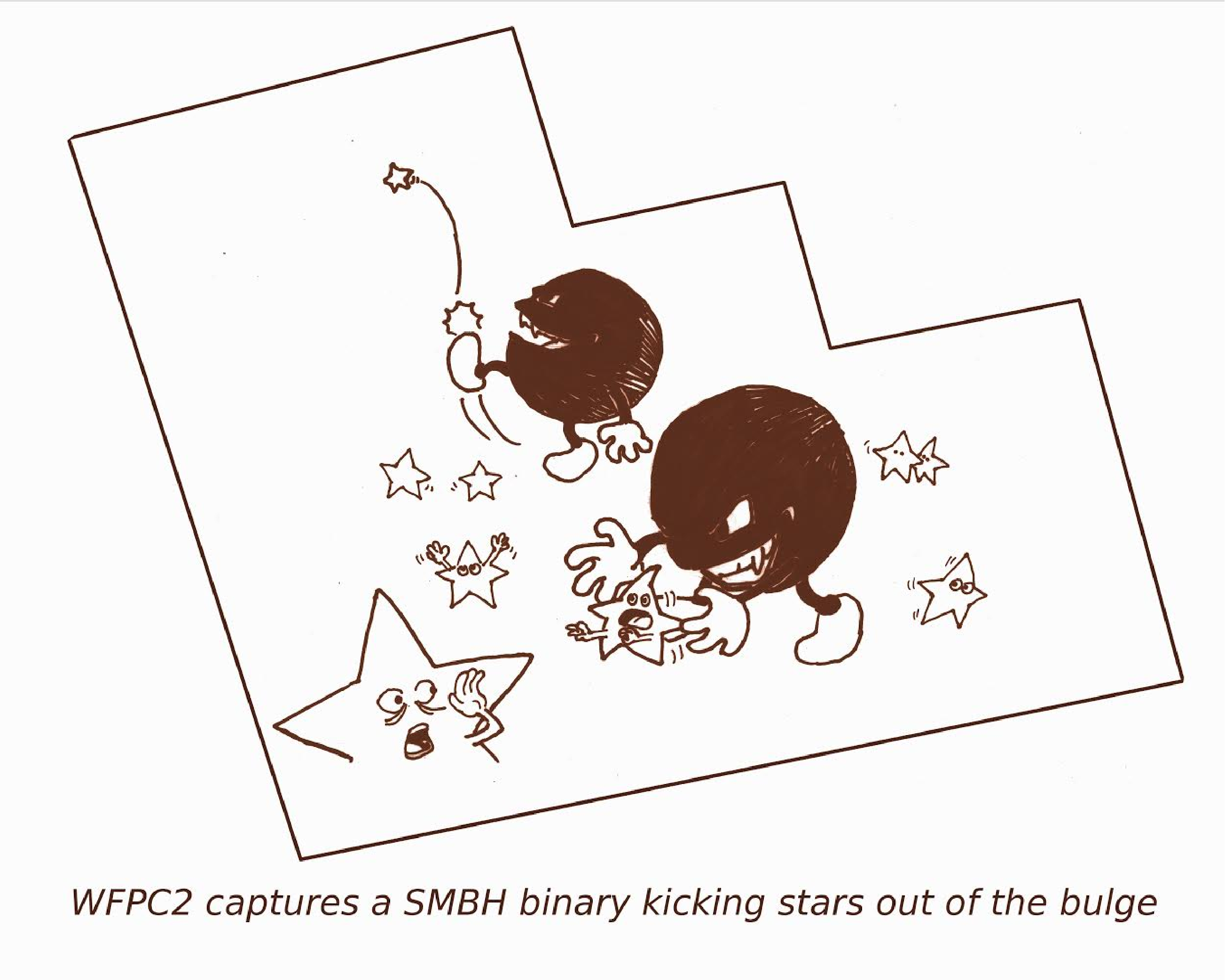}
\caption{Cartoon showing a pair of supermassive black holes kicking stars away
as they dance towards coalescence at the centre of a galaxy.  Credit: Paolo Bonfini.}
\label{Fig7}
\end{figure}

Using the core-S\'ersic model to quantify the central flux deficit, and in
turn the stellar mass deficit, Graham (2004) discovered $M_{\rm def} \approx 2
M_{\rm bh}$.  Previously it was thought that $M_{\rm def}/M_{\rm bh}$ was, on
average, an order of magnitude greater (e.g.\ Milosavljevi{\'c} et al.\ 2002;
Ravindranath, Ho \& Filippenko 2002), which required a troublingly large
number of merger events given that the ejected mass should roughly scale with
$N\,M_{\rm bh}$, where $N$ is the cumulative number of (equivalent major) dry
merger events (Milosavljevi{\'c} \& Merritt 2001; Merritt 2006a).  Using the
core-S\'ersic model, these new lower mass ratios were also found by Ferrarese
et al.\ (2006a) and Hyde et al.\ (2008).  Using the idea from Graham et
al.\ (2003) that cores can be measured as a deficit of light relative to the
inward extrapolation of the outer S\'ersic profile, but fitting the S\'ersic
model rather than core-S\'ersic model and identifying the sizes of depleted
cores by eye, Kormendy \& Bender (2009) reported notably larger mass ratios
(typically close to 10 or higher).  Hopkins \& Hernquist (2010) subsequently
resolved this issue in a model-independent manner and revealed that the
core-S\'ersic model measurements of the central mass deficits were correct.
Most recently, Rusli et al.\ (2013b) found that $\sim$80\% of their 23
galaxies have $1 < M_{\rm def}/M_{\rm bh} < 5$, while Dullo \& Graham (2014)
reported typical values for their sample of 31 galaxies to be $0.5 < M_{\rm
  def}/M_{\rm bh} < 4$.

Although the central mass deficit and break radius are obviously not
fundamental parameters in establishing the spheroid-(black hole) connection
--- simply because many galaxies have black holes but not partially depleted
cores --- there is nonetheless an $M_{\rm bh}$--$R_b$ relation (Lauer et
al.\ 2007)\footnote{Lauer et al.\ (2007) found that using the radius where
  the negative, logarithmic slope of the surface brightness profile equals 0.5
  (which matches well with the core-S\'ersic break radius: Dullo \& Graham
  2012, their section~5.2) produces a stronger relation than obtained when
  using the Nuker model break radii.} and an $M_{\rm bh}$--$M_{\rm def}$
relation (e.g.\ Graham 2004; Rusli et al.\ 2013b; Dullo \& Graham 2014).
This relation simply exists over a restricted mass range.  
Dullo \& Graham (2014, their Eq.~18) reported that $M_{\rm def} \propto M_{\rm
  bh}^{3.70\pm0.76}$ for the population ensemble (not to be confused with
growth in individual galaxies).  This is of interest for several reasons.  One
of which is that it may provide insight into the merging scenario, which
currently has an unresolved problem.  In general, galaxies with the greatest
$M_{\rm def}/M_{\rm bh}$ ratio should have experienced the highest number of
major dry mergers, and due to the increase in black hole mass but stagnation
in velocity dispersion associated with such mergers (e.g.\ Ostriker \& Hausman
1977; Hausman \& Ostriker 1978; Ciotti \& van
Albada 2001), they should be offset to high black hole masses in the $M_{\rm
  bh}$--$\sigma$ diagram (see Volonteri \& Ciotti 2013). However, they are not
(Savorgnan \& Graham 2015).

Within low-luminosity early-type galaxies, the nuclear star cluster can be
slightly offset ($\sim$100 parsec) from the galaxy's photometric centre
(Binggeli et al.\ 2000; Barazza et al.\ 2003).  This is thought to be due to
the dense star cluster's harmonic oscillation within the weak gravitational
gradient of the galaxy's core.  The amplitude of the nuclear cluster's rocking
back and forth motion is expected to be greater in spheroids with lower
S\'ersic index, because they have lower central stellar densities and
shallower inner density profiles, and thus less well defined gravitational
centres over a greater fraction of their half-light radii (see Terzi\'c \&
Graham 2005, their figure~2).  Similarly, high-luminosity core-S\'ersic
spheroids have somewhat weakened gravitational centres (Terzi\'c \& Graham
2005, their figure~3) due to the partial depletion of stars in their cores.
One may then expect to find the supermassive black holes slightly offset from
the photometric centres of core-S\'ersic galaxies (Miller \& Smith 1992; Taga
\& Iye 1998).  However a mechanism capable of creating more extreme ($> 1$
kpc) offsets is the recoil from the emission of anisotropic gravitational
radiation that a newly merged black hole may receive (e.g.\ Bonnor \&
Rotenberg 1961; Peres 1962; Bekenstein 1973).  The linear momentum carried
away by the gravitational wave is balanced by a kick imparted to the black
hole.  This recoil process has the ability to evacuate a much greater loss
cone, and has been proposed as an explanation for some cores having large
$M_{\rm def}/M_{\rm bh}$ ratios (e.g.\ Boylan-Kolchin et al.\ 2004; Campanelli
et al.\ 2007; Gualandris \& Merritt 2008, 2012), which have been observed in
NGC~1399 and NGC~5061.  While only small spatial offsets are known for black
holes in galaxies with directly measured black hole masses (e.g.\ Batcheldor
et al.\ 2010; Lena et al.\ 2014), if this process is operating one might
expect to see greater displacements (e.g.\ Blecha et al.\ 2012) of black holes
in galaxies with larger $M_{\rm def}/M_{\rm bh}$ ratios.  However, if the
damping timescale of the recoil-induced oscillation is sufficiently short, one
may not find this correlation.

In passing, it might be remiss if a few words were not said about the
gravitational wave signals expected from the final coalescence of massive
black holes after they have scoured out the cores of massive spheroids,
preferentially removing stars on plunging radial orbits (e.g.\ Quinlan \&
Hernquist 1997; Milosavljevi\'c \& Merritt 2001; Thomas et al.\ 2014).  Binary
AGN, and thus massive black holes, are now known in several galaxies
(e.g.\ Komossa et al.\ 2003; Liu et al.\ 2014, and references therein).  The
rapidly changing gravitational field as the black holes spiral (and thus
accelerate) around each other, generates a gravitational wave-like ripple
which radiates out into space (e.g.\ Buonanno \& Damour 2000; Barack \& Cutler
2004; Baker et al.\ 2006; Blanchet 2006; Sesana 2010; Amaro-Seoane et
al.\ 2012).  Travelling at the speed of light, the amplitude of the wave
decays linearly (rather than quadratically) with distance and, also unlike
light, passes unimpeded through both space and matter.  Due to the large
orbital size of the binary black hole, space-based interferometers at great
separations are required to sample the long wavelength of the waves generated
by the black hole binary.  Building on the hopes of the Laser Interferometer
Space Antenna (LISA: Danzmann \& R\"udiger 2003), the European LISA Pathfinder
mission\footnote{\url{http://sci.esa.int/lisa-pathfinder/}} (LPF: Anza et
al.\ 2005; McNamara 2013), formerly known as SMART-2, offers the very exciting
promise of detecting these waves predicted by Einstein's theory of relativity
but not yet observed (Will 2006).

\section{Intermediate mass black holes and the (black hole)--(nuclear cluster)
  connection} 

\label{Sec_bh-nc}

As was noted in section~\ref{SubSec_bend}, the bent $M_{\rm bh}$--$M_{\rm
  sph}$ relation offers hope for detecting the missing population of
intermediate mass black holes.  This is because the linear $M_{\rm
  bh}$--$M_{\rm sph}$ relation predicts $10^2 < M_{\rm bh}/M_{\odot} < 10^5$
black hole masses in smaller / fainter spheroids.  Although we may not have
the spatial resolution at optical/near-infrared wavelengths to resolve the
sphere-of-influence of these black holes, and thus directly measure their
masses from Keplerian kinematics, there is an independent method which can be
used to predict (strengthen / reject) the likely existence of such
intermediate mass black holes.  It is based on the observation that the black
hole mass correlates with the AGN radio and X-ray flux in such a way that they
define a 2-dimensional surface in 3-parameter space, which has been dubbed the
`fundamental plane of black hole activity' (Merloni et al.\ 2003).  Therefore,
obtaining radio and X-ray data is expected to prove fruitful in the hunt for
the elusive intermediate mass black holes.  Preferably, this data should be
obtained simultaneously because the AGN are known to vary in their flux output
over timescales of days.

One of the best candidates for an intermediate mass black hole is the
ultraluminous X-ray source HLX-1 in the galaxy ESO 243–49 (Farrell et
al.\ 2009; Webb et al.\ 2014).  Interestingly, this 9,000 solar mass black
hole candidate does not reside near the centre of its host galaxy but in a
compact star cluster (Soria et al.\ 2010; Wiersema et al.\ 2010; Farrell et
al.\ 2012) located at a projected distance of $\sim$3 kpc from the galaxy's
nucleus, perhaps shedding insight into the formation location of intermediate
mass black holes (see also Mezcua et al.\ 2013, 2015 in regard to an off-centered
intermediate mass black hole candidate in NGC~2276). 
Despite early hopes for intermediate mass black holes in
globular clusters (e.g.\ Gerssen et al.\ 2003; Gebhardt et al.\ 2005; Noyola
et al.\ 2010; L{\"u}tzgendorf et al.\ 2013, and references therein), there are
not yet any definite candidates (e.g.\ van den Bosch et al.\ 2006; Hurley
2007; Anderson \& van der Marel 2010; Vesperini \& Trenti 2010; Lanzoni et
al.\ 2013; Lanzoni 2015). Observational research programs (e.g.\ Bellini et
al.\ 2014; Lapenna et al.\ 2014) continue the hunt as the formation of
intermediate mass black holes in dense star clusters seems probable
(e.g.\ Miller \& Hamilton 2002; Baumgardt et al.\ 2004; G\"urkan et al.\ 2004;
Portegies Zwart et al.\ 2004).

\begin{figure}
\includegraphics[scale=0.47, angle=-90]{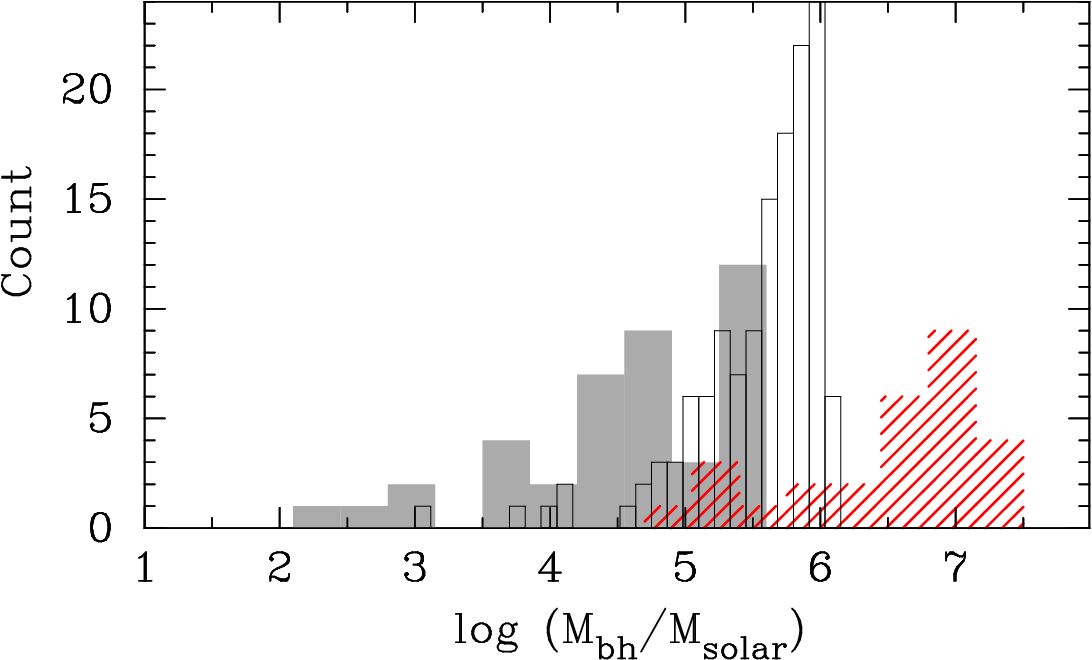}
\caption{Predicted black hole masses. 
The solid histogram was obtained using the $M_{\rm bh}$--$L_K$ relation for S\'ersic
spheroids applied to the 
$K$-band bulge magnitudes in Graham \& Scott (2013, their table~6). 
The open histogram was obtained using the $M_{\rm bh}$--$M_{\rm sph}$ relation
for S\'ersic spheroids (shown in Figure~\ref{Fig2}) applied to the dwarf galaxy
masses in Reines et al.\ (2013, their table~1). 
The shaded histogram was obtained in the same way but using the dwarf galaxy stellar 
masses in Moran et al.\ (2014, their table~1).  The fainter bulges are
expected to contain the least massive black holes. }
\label{Fig8}
\end{figure}

Aside from globular clusters, some of the dense star clusters found in the
nuclei of many low- and intermediate-luminosity spheroids (e.g.\ Reaves 1983;
Binggeli et al.\ 1985; Phillips et al.\ 1996; Carollo et al.\ 1997) are
already known to house massive black holes.  Ferrarese et al.\ (2006b) and
Wehner \& Harris (2006) originally suggested that these star clusters may be
the low-mass extension of the supermassive black holes, in the sense that
galaxies housed one type of nucleus or the other.  However this idea was soon
modified when it was realised that such clusters and massive black hole
coexist in substantial numbers of galaxies (e.g.\ Gonz{\'a}lez Delgado et
al.\ 2008; Seth et al.\ 2008; Graham \& Spitler 2009).  Ongoing efforts
have revealed that nuclear star clusters do not follow the same mass scaling
relations as supermassive black holes (Graham 2012b; Leigh et al.\ 2012;
Neumayer \& Walcher 2012; Scott \& Graham 2013), and the search for intermediate
mass black holes continues.  Among the most promising targets are the low mass
bulges of disc galaxies hosting an AGN (Graham \& Scott 2013) and the low mass
dwarf galaxies which also display AGN activity (e.g.\ Reines et al.\ 2013;
Moran et al.\ 2014); see Figure~\ref{Fig8}. 

Just as there is a relation between spheroid luminosity and the central
surface brightness\footnote{Technically it is the central surface brightness
  of the spheroid excluding blips from additional nuclear components such as
  star clusters.}  of the 
spheroid --- until the onset of partially depleted cores in massive spheroids
--- there is also a relationship between spheroid luminosity and the
brightness of the nuclear star clusters that they host (Balcells et al.\ 2003;
Graham \& Guzm\'an 2003).  In a somewhat similar manner to the establishment
of the $M_{\rm bh}$--$\mu_0$ relation presented in Section~\ref{Sec_mu0}, one
can predict what the $M_{\rm bh}$-$M_{\rm nc}$ relation should be like.
Graham (2015a) combined the relation 
$M_{\rm bh} \propto M_{\rm sph}^2$ for the S\'ersic spheroids
(Section~\ref{SubSec_bend}) with the relation 
$M_{\rm nc} \propto M_{\rm sph}^{0.6–-1.0}$ (references above)
to obtain 
$M_{\rm bh} \propto M_{\rm nc}^{2–-3.3}$.
A consistent result was obtained by coupling the relation 
$M_{\rm bh} \propto \sigma^{5.5}$ (Section~\ref{Sec_M-sigma}) with
$M_{\rm nc} \propto \sigma^{1.6–-2.7}$ (references above)
to give 
$M_{\rm bh} \propto M_{\rm nc}^{2.0–-3.4}$.  
Massive black holes therefore grow rapidly within their host star cluster,
until it is evaporated (e.g.\ Bekki \& Graham 2010) or partially devoured
(e.g.\ Hills 1975; Frank \& Rees 1976; Murphy et al.\ 1991; Komossa 2013;
Donato et al.\ 2014; Vasiliev 2014).  However, disentangling which came first
may be an interesting pursuit, and just as there are different types of
bulges, there may be different types of nuclear star clusters (e.g.\ Turner et
al.\ 2012).  This $M_{\rm bh}$-$M_{\rm nc}$ relation is somewhat complementary to the 
$M_{\rm bh}$-$M_{\rm def}$ relation, with each applicable at opposing ends of
the black hole mass range currently accessible. Such co-occupancy of black
holes and nuclear star clusters is a likely source of stellar tidal disruption
events (Komossa et al.\ 2009, 2013 and references therein) and gravitational
wave emission from the inspiralling of compact stellar remnants
(e.g.\ Hils \& Bender 1995; 
Amaro-Seoane et al.\ 2007 and references therein), predictions for
which are dramatically modified when using the new, near-quadratic $M_{\rm
  bh}$-$M_{\rm sph}$ relation (Mapelli et al.\ 2012).  Further quantifying the
coexistence of massive black holes in dense, compact, nuclear star clusters
should help us to predict the occurrence of, and better understand, these
exciting phenomenon.

\section{The $M_{\rm bh}$--$M_{\rm halo}$ relation} 
\label{Sec_halo}

Ferrarese et al.\ (2002) have revealed that there is a relationship
between the black hole mass and the galaxy halo mass (baryons plus dark
matter), as traced by the circular velocity at large radii (used as a proxy
for the halo's virial radius).  Due to the relation between this rotational
velocity and the galaxy's velocity dispersion (see also Baes et al.\ 2003;
Pizzella et al.\ 2005; Ferrarese \& Ford 2005, their Eq.~21)\footnote{It
  should be noted that the dynamical study by Kronawitter et al.\ (2000) and
  Gerhard et al.\ (2001), which led to the relationship between the circular
  velocity and the velocity dispersion for elliptical galaxies, was based on a sample
  of elliptical galaxies that had very similar absolute
  magnitudes. Consequently, these galaxies will have similar structural and
  dynamical profiles, and thus their $v_{\rm circ}$--$\sigma$ relationship may
  not be applicable to lower- or higher-luminosity elliptical galaxies with
  different S\'ersic indices, i.e.\ concentration, and dynamical profiles
  (e.g.\ Ciotti 1991).}  one can expect an $M_{\rm bh}$--$M_{\rm halo}$
relation.  The extent of this relationship may be applicable only to galaxies with
large bulges (or $v_{\rm circ} > \sim100$ km s$^{-1}$ or $\sigma > \sim100$ km
s$^{-1}$), because of the breakdown in the relationship between circular
velocity and velocity dispersion for lower mass systems (e.g.\ Zasov et al.\ 2005; 
Ho 2007; Courteau et al.\ 2007). 
Nonetheless, this would make the relationship exist over a larger mass range
than the $M_{\rm bh}$--$R_{\rm b}$ and $M_{\rm bh}$--$M_{\rm def}$ relations
(Section~\ref{Sec_core}).

For galaxies built from major dry merger events, in which the black hole mass
and the galaxy stellar mass simply add together, the dark matter must also add
in this linear fashion.  This would then establish a linear $M_{\rm
  bh}$--$M_{\rm halo}$ relation --- just as there is a linear $M_{\rm
  bh}$--$M_{\rm sph}$ relation preserving the $M_{\rm bh}/M_{\rm sph}$ ratio
--- at high masses ($M_{\rm bh} > \sim 10^8 \, M_{\odot}$).  This appears to
be consistent with the data in Ferrarese et al.\ (2002, their figure~5).
However, their linear regression to the fuller sample gives $M_{\rm bh}
\propto M_{\rm halo}^{1.65 \, - \, 1.82}$, which is in remarkable agreement
with the prediction $M_{\rm bh} \propto M_{\rm halo}^{5/3}$ by Haehnelt,
Natarajan \& Rees (1998).  Although, with a different sample, Baes et
al.\ (2003) reported $M_{\rm bh} \propto M_{\rm halo}^{1.27}$.  Curiously, for
elliptical galaxies not built from dry mergers\footnote{Equal mass, 
  (major) dry mergers preserve the $M_{\rm halo}/L$ ratio and therefore
  galaxies built from major dry mergers follow the sequence $M_{\rm halo}/L
  \propto L^0$.}, the prediction by Haehnelt et al.\ (1998) transforms into
$M_{\rm bh} \propto L^{20/9}_{\rm gal} (= L^{2.22}_{\rm gal})$ if $M_{\rm
  halo}/L_{\rm gal} \propto L^{1/3}_{\rm gal}$ (J{\o}rgensen et al.\ 1996;
Cappellari et al.\ 2006). This near-quadratic relation has been seen before in
Section~\ref{Sec_M-M}.

\subsection{Globular cluster systems}

Lending support to the $M_{\rm bh}$--$M_{\rm halo}$ relation is the connection
between black hole mass and the halo of globular clusters that swarm around
galaxies, both in terms of their number (Burkert \& Tremaine 2010; Harris \&
Harris 2011; Rhode 2012; Harris et al.\ 2014) and their velocity dispersion
(Sadoun \& Colin 2012; Pota et al.\ 2013).  In Burkert \& Tremaine (2010) they
used a (self-admittedly limited) sample of 13 galaxies for which the black
hole mass and the number of globular clusters was known. They observed an rms
scatter of just 0.21 dex about their optimal relation in the $\log(M_{\rm
  bh})$ mass direction.  Not surprisingly this attracted some interest 
(e.g.\ Snyder et al.\ 2011) because 
it was half of the value observed in the $M_{\rm bh}$--$\sigma$ diagram.
However as more galaxies have been added, the scatter about the relation
involving the globular clusters has increased. 

The globular cluster system around individual galaxies are known to display a
bimodality in their colour, with the red (metal rich) globular clusters
thought to be associated with the galaxy's bulge while the blue (metal-poor)
globular clusters are thought to be connected with the halo (Ashman \& Zepf
1992; Forbes et al.\ 1997).  Using both the observed velocity dispersion of
the globular cluster system, and the velocity dispersion with the rotational
component of the system subtracted, Pota et al.\ (2013) report that while a
correlation with black hole mass is evident, it is not yet clear if the black
hole mass is better correlated with the red (bulge) or the blue (halo)
globular cluster sub-population.

\section{The $M_{\rm bh}$--(spiral arm pitch angle) connection}
\label{Sec_pitch}

While the applicability of the $M_{\rm bh}$--$M_{\rm halo}$ relation in lower
mass spiral galaxies is unclear, there is a somewhat complementary relation
which only operates in spiral galaxies. Seigar et al.\ (2008; see also
Ringermacher \& Mead 2009; Treuthardt et al.\ 2012 and Berrier et al.\ 2013)
have presented the relation between black hole mass and spiral arm pitch
angle.  The spiral arm pitch angle (e.g.\ Puerari et al.\ 2014, and references
therein) is of course known to vary along the
Hubble-Jeans sequence, as does the bulge-to-total flux ratio, or more correctly the
luminosity of the bulge (e.g.\ Yoshizawa \& Wakamatsu 1975; Ostriker 1977;
Meisels \& Ostriker 1984; Trujillo et al.\ 2002), which may explain the 
black hole connection with the pitch angle.  As with the
radial concentration of the bulge light, the pitch angle has the advantage
that it can be measured from
photometrically uncalibrated images and therefore offers an easy means to
predict black hole masses (perhaps even when there is no bulge\footnote{The
  $M_{\rm bh}$--(pitch angle) relation is yet to be established for a sample
  of bulgeless galaxies.}), from which one can then do clever things like determine
the black hole mass function in spiral galaxies (Davis et al.\ 2014).

Given that this is obviously a secondary relation, although the low level of
scatter reported by Davis et al.\ (2014) is intriguing, less shall be said 
about this than the relations involving a spheroid's central concentration and density
of stars (Sections~\ref{Sec_n} and \ref{Sec_mu0}). 

\section{Fundamental planes: adding a third parameter} 
\label{Sec_fun}

As noted in Section~\ref{Sec_bh-nc}, stellar and supermassive black holes
roughly define a plane within the 3-dimensional space of black hole mass,
radio power and X-ray luminosity (Merloni et al.\ 2003; Heinz \& Sunyaev 2003;
Falcke et al.\ 2004; K\"ording et al.\ 2006; Li et al.\ 2008).  While this is
both interesting in its own right and highly useful, the relationship between the black
hole mass, accretion disc and jet is of a different nature to the other
relations presented in this article and as such is not detailed here as it is
an AGN phenomenon.

One of the early attempts to introduce a third parameter into the (black
hole)--(host galaxy) scaling relations was by Marconi \& Hunt (2003).  They
used the effective half light radius ($R_{\rm e}$) of the spheroid, together
with the velocity dispersion ($\sigma$), to derive a rough virial mass for the
spheroid ($M_{\rm virial} \propto \sigma^2\, R_{\rm e}$).  They found that the
total vertical scatter about their $M_{\rm bh}$--$M_{\rm virial}$ relation was slightly less
than that about their $M_{\rm bh}$--$\sigma$ relation (0.25 dex vs 0.30 dex). 
Using a sample of elliptical galaxies, Feoli \& Mele (2005; see also 
Feoli \& Mancini 2009, 2011) 
reported on a black hole mass relation with the kinetic energy of the host
galaxy such that $M_{\rm bh} \propto (M_{\rm gal}\sigma^2)^{\alpha}$, where
$0.87 < \alpha < 1.00$ and $M_{\rm gal}$ was derived assuming $R^{1/4}$ light
profiles\footnote{It should be noted that the assumption of $R^{1/4}$ light
profiles can introduce a systematic bias with galaxy mass, S\'ersic index and
effective radius (e.g.\ Trujillo et al.\ 2001; Brown et al.\ 2003).}. 
Given that $M_{\rm gal}$ roughly scales as 
$\sigma^2\, R_{\rm e}$, their kinetic energy expression roughly scales with 
$\sigma^4\, R_{\rm e}$. 
Additional variations of this theme, searching for a fundamental plane
using combinations of $\sigma$ and $R_{\rm e}$ can be found in 
de Francesco et al.\ (2006), who effectively suggested independent exponents for
$\sigma$ and $R_{\rm e}$, in Aller \& Richstone (2007) in terms of the
gravitational binding energy, and in Hopkins et al.\ (2007) and Soker \& Meiron
(2011). 
Given the existence of {\it the} Fundamental Plane (Djorgovski \& Davis 1985)
linking the velocity dispersion with the mean effective surface brightness
($\langle \mu \rangle_{\rm e}$) and effective half light radius, the presence
of the $M_{\rm bh}$--$\sigma$ relation additionally suggests that there should be an
$M_{\rm bh}$--$(\langle \mu \rangle_{\rm e}, R_{\rm e})$ plane (Barway 
\& Kembhavi 2007). 

With all of these attempts to define different planes, 
there are two issues that require attention: (i) barred galaxies, and (ii) the
accuracy\footnote{It could be argued that a third issue is the accuracy of the
  black hole masses (Merritt 2013).} and thus usefulness of $R_{\rm e}$.

First, the increased scatter in the $M_{\rm bh}$--$\sigma$ diagram due to the
inclusion of barred galaxies was reported by Graham (2008a,b) and Hu (2008).
Moreover, Graham (2008a) showed that once the barred galaxies were removed, 
there was no reduction in scatter when going from the $M_{\rm bh}$--$\sigma$
diagram to the $M_{\rm bh}$--$(\sigma, R_{\rm e})$ diagram.  If there is a 
more fundamental relation with some combination of $\sigma$ and $R_{\rm e}$,
than compared with $\sigma$ alone, this should not have been observed.  
The simulations of Younger et al.\ (2008, their Fig.9) show that 
(merger built) classical bulges follow a plane, without the need to include 
(secular-disk-evolution built) pseudobulges. 
Therefore, if the lower scatter about the hybrid relations is only achieved
when including the barred galaxies, it suggests that something else is
responsible for the reduction, such as barred galaxies having smaller $R_{\rm
  e}$ values than the elliptical galaxies which dominate at the high mass end
of one's sample.  Younger et al.\ (2008) suggested that the 
relatively small dynamic range among the non-barred galaxies with direct 
black hole mass measurements may have been inadequate to provide a significant
detection of this third parameter and thus a plane. 
It would be interesting to repeat the tests which 
searched for an optimal plane among the non-barred galaxies, 
but now using the larger galaxy samples which are available. 
However this brings us to the second issue.

\begin{figure}[t]
\includegraphics[scale=.46]{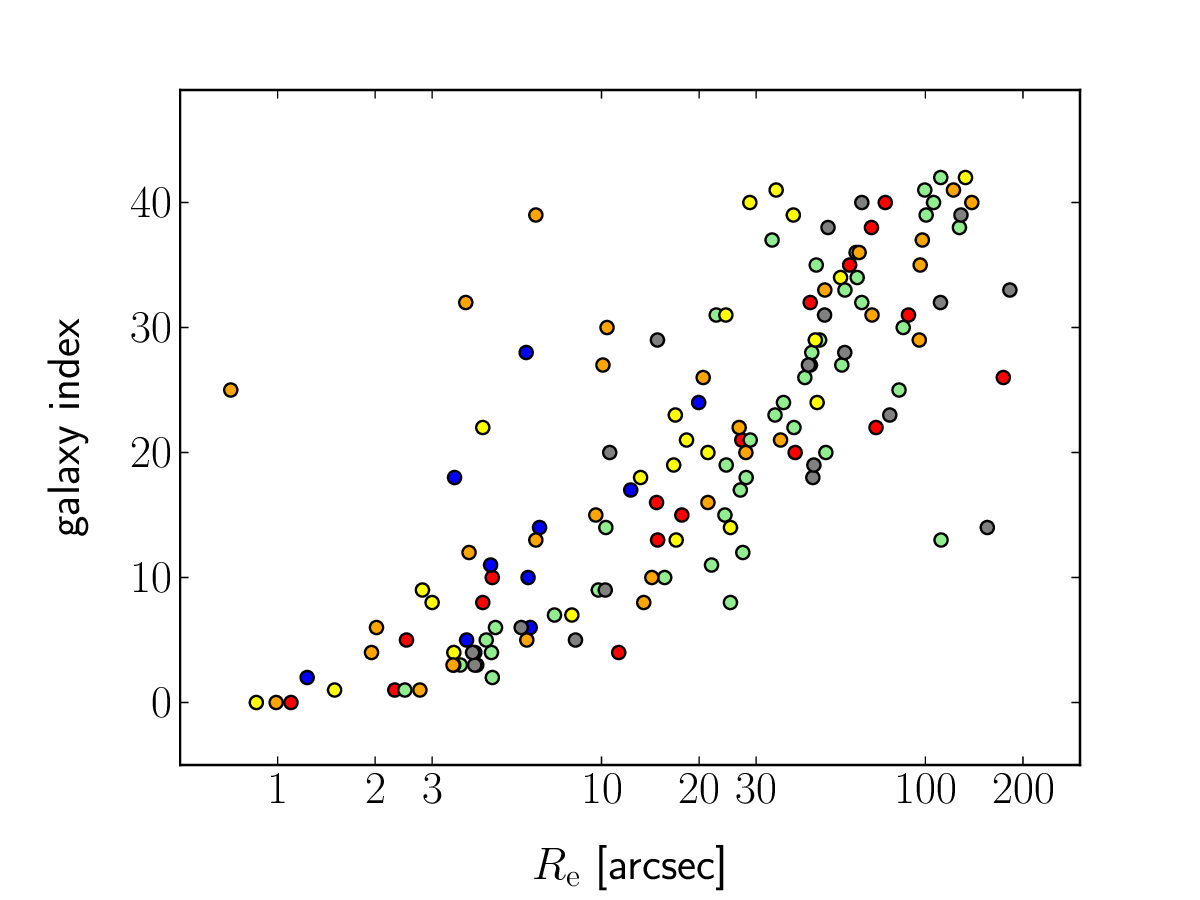}
\caption{Major-axis effective half-light radii $R_{\rm e}$ 
for the spheroidal component of 43 galaxies (having directly 
measured black hole masses) as determined by different authors. 
Figure taken from Savorgnan \& Graham (2016a).  
Legend: 
red = Graham \& Driver (2007); 
blue = Laurikainen et al.\ (2010);
green = Sani et al.\ (2011); 
yellow = Vika et al.\ (2012); 
gray = Beifiori et al.\ (2012); 
orange = L\"asker et al.\ (2014a).
}
\label{Fig9}
\end{figure}

Given that there have been errors in the measurement of the S\'ersic indices
$n$ (as revealed by Savorgnan et al.\ 2013), there are thus errors in the
measurements of the published, effective half light radii $R_{\rm e}$ (see also
Bernardi et al.\ 2014).  Harris 
et al.\ (2014) show the large range of $R_{\rm e}$ values (for the same
spheroid) reported by different authors for spheroids with directly measured
black hole masses.  A similar plot is shown in Figure~\ref{Fig9} but this time
restricting the data to that obtained from S\'ersic $R^{1/n}$ model fits by
different authors.  Consequently, attempts to use $R_{\rm e}$ for measuring
dynamical masses ($\propto \sigma^2\, R_{\rm e}$) or as a third parameter to
mop up some of the scatter about the $M_{\rm bh}$--$\sigma$ relation should at
this time be treated with caution.

\section{Concluding remarks}
\label{Sec_sum}

The ``attraction'' of black holes is vast, as evinced by a huge literature on
the subject, of which but a small fraction is noted here.  The fundamental
physical connection between black-hole and bulge growth still awaits
discovery.  While it is expected that we may narrow in on the solution as we
keep plugging away at more black hole mass measurements, coupled with
improving the accuracy of all quantities involved, it is reasonable to expect 
that something unexpected may be discovered, such is the nature and joy of our
collective pursuit.

Given the role that pulsars played in convincing the community that black
holes may exist in 1967--8, it is perhaps fitting that arrays of pulsar
beacons are used today (e.g.\ Sesana et al.\ 2008; Hobbs et al.\ 2010 ; Kramer
\& Champion 2013) to try and detect the bob and sway of the space antennae as
anticipated gravitational waves --- from the inspiral of supermassive black
holes at the centres of newly merged galaxies --- wash by oblivious to our
solar system.  The future direct detection of such gravitational radiation would
provide another strong test of Einstein's theory of general relativity
(e.g.\ Will 2006, 2014), which, starting 100 years ago, led to the modern
prediction of dense, dark stars and supermassive black holes.

\acknowledgements 
This review was made possible by Australian Research Council funding through
grant FT110100263. 
This research has made use of NASA's Astrophysics Data System Bibliographic
Services, and the NASA/IPAC Extragalactic Database (NED).

\end{document}